\documentclass[robotics,article,accept,moreauthors,pdftex,10pt,a4paper]{mdpi} 
\firstpage{1} 
\pubvolume{8}
\issuenum{2}
\articlenumber{5}
\pubyear{2019}
\copyrightyear{2019}
\history{Received: 7 July 2019; Accepted: 19 September 2019; Published: date}

\usepackage{environ}
\NewEnviron{myequation}{%
\begin{equation}
\scalebox{0.88}{$\BODY$}
\end{equation}
}

\usepackage{amssymb,amsmath,amstext,amsfonts}
\usepackage{mathtools}
\usepackage{bm}

\usepackage{xcolor}
\usepackage{booktabs}
\usepackage{url}

\usepackage{graphicx}
\usepackage{float}

\usepackage{algorithm,algpseudocode,float,algcompatible,amsmath}
\newlength\hrulethickness
\makeatletter
\renewcommand\fs@ruled{\def\@fs@cfont{\bfseries}\let\@fs@capt\floatc@ruled
  \def\@fs@pre{\hrule height \hrulethickness depth0pt \kern2pt}%
  \def\@fs@post{\kern2pt\hrule height \hrulethickness depth0pt \relax}%
  \def\@fs@mid{\kern2pt\hrule height \arrayrulewidth depth0pt \kern2pt}%
  \let\@fs@iftopcapt\iftrue}
\makeatother

\usepackage{subfigure} 
\makeatletter
\renewcommand{\@thesubfigure}{\normalsize(\textbf{\alph{subfigure}})}
\makeatother

\usepackage{siunitx}
\usepackage{nomencl} 
\makenomenclature

\usepackage{xcolor}
\definecolor{darkbrown}{rgb}{0.4, 0.26, 0.13}
\newcommand\crule[3][darkbrown]{\textcolor{#1}{\rule{#2}{#3}}}

\usepackage{wasysym}

\usepackage{amssymb}
\usepackage{amsmath}
\usepackage{graphicx}
\usepackage{pifont}
\DeclareRobustCommand*{\pmstar}{%
	\text{%
		\resizebox{!}{.5\height}{\ding{107}}%
	}%
}

\Title{Online Multi-Objective Model-Independent Adaptive Tracking Mechanism for Dynamical~Systems}

\Author{Mohammed Abouheaf $^{1,2}$\orcidA{}, Wail Gueaieb $^{1,}$*\orcidB{} and Davide Spinello $^{3}$\orcidC{}}

\AuthorNames{Mohammed Abouheaf and Wail Gueaieb and Davide Spinello}

\address{$^{1}$ \quad School of Electrical Engineering and Computer Science, Faculty of Engineering, University of Ottawa, {Ottawa, ON~K1N~6N5,}~Canada; 
\\$^{2}$ \quad Department of Electrical Engineering, College of Energy Engineering, Aswan University, Aswan, 81521, Egypt
\\$^{3}$ \quad Department of Mechanical Engineering, Faculty of Engineering, University of Ottawa, {Ottawa, ON~K1N~6N5,}~Canada;}
  
\corres{Correspondence: w{g}ueaieb@uottawa.ca; Tel.: +1-613-562-5800 (ext. 2158)}

\abstract{
The optimal tracking problem is addressed in the robotics literature by using a variety of robust and adaptive control approaches. However, these schemes are associated with implementation limitations such as applicability in uncertain dynamical environments with complete or partial model-based control structures, complexity and integrity in discrete-time environments, and~scalability in complex coupled dynamical systems. An online adaptive learning mechanism is developed to tackle the above limitations and provide a generalized solution platform for a class of tracking control problems. This scheme minimizes the tracking errors and optimizes the overall dynamical behavior using simultaneous linear feedback control strategies. Reinforcement learning approaches based on value iteration processes are adopted to solve the underlying Bellman optimality equations. The resulting control strategies are updated in real time in an interactive manner without requiring any information about the dynamics of the underlying systems. Means of adaptive critics are employed to approximate the optimal solving value functions and the associated control strategies in real time. The proposed adaptive tracking mechanism is illustrated in simulation to control a flexible wing aircraft under uncertain aerodynamic learning environment.
}


\begin{document}

\section{Introduction}
	
Adaptive tracking control algorithms employ challenging and complex control architectures under prescribed constraints about the dynamical system parameters, initial tracking errors, and~stability conditions~\cite{JIANG1997,Tseng2001}. These schemes may include cascade linear stages or over-parameterize the state feedback control laws to solve the tracking problems~\cite{Lefeber2003,ZHAO2015}. Among~the challenges associated with this class of control algorithms, is the need to have full or partial knowledge of the dynamics of the underlying systems, which can degrade their operation in the presence of uncertainties~\cite{Kamalapurkar2017,Zhang2016}. Some approaches employ tracking error-based control laws and cannot guarantee overall optimized dynamical performance. This motivated the introduction of flexible innovative machine learning tools to tackle some of the above limitations. In~this work, online value iteration processes are employed to solve optimal tracking control problems. The~associated temporal difference equations are arranged to optimize the tracking efforts as well as the overall dynamical performance. Linear~quadratic utility functions, which are used to evaluate the above optimization objectives, result~in two model-free linear feedback control laws which are adapted simultaneously in real time. The~first feedback control law is flexible to the tracking error combinations (i.e., possible higher-order tracking error control structures compared to the traditional continuous-time Proportional-Derivative (PD) or Proportional-Integral-Derivative (PID) control mechanisms), while the second is a state feedback control law that is designed to obtain an optimized overall dynamical performance, while affecting the closed-loop characteristics of the system under consideration. This~learning approach does not over-parameterize the state feedback control law and it is applicable to uncertain dynamical learning environments. The~resulting state feedback control laws are flexible and adaptable to observe a subset of the dynamical variables or states, which is really convenient in cases where it is either hard or expensive to have all dynamical variables measured. Due~to the straightforward adaptation laws, the~tracking scheme can be employed in systems with coupled  dynamical structures. Finally, the~proposed method can be applied to nonlinear systems, with~no requirement of output feedback~linearization.

To showcase the concept in hand and to highlight its effectiveness under different modes of operation, a~trajectory-tracking system is simulated using the proposed machine learning mechanism for a flexible wing aircraft. Flexible wing systems are described as two-mass systems interacting through kinematic constraints at the connection point between the wing system and the pilot/fuselage system (i.e., the hang-strap point)~\cite{Kilkenny_1983,Kilkenny_1984,Kilkenny_1986,Blake_1991}. The~modeling approaches of the flexible wing aircraft typically rely on finding the equations of motion using perturbation techniques~\cite{Kroo_1983}. The resulting model decouples the aerodynamics according to the directions of motion into the longitudinal and lateral frames~\cite{Spottiswoode_2001}. Modeling this type of aircraft is particularly challenging due to the time-dependent deformations of the wing structure even in steady flight conditions~\cite{Cook2006,Cook_Kilkenny_1986,DE_MATTEIS_1990,De_Matteis_1991}. Consequently, model-based control schemes typically degrade the operation under uncertain dynamical environments. The~flexible wing aircraft employs weight shift mechanism to control the orientations of the wing with respect to the pilot/fuselage system. Thus, the~aircraft pitch/roll orientations are controlled by adjusting the relative centers of gravities of these highly coupled and interacting systems~\cite{Kilkenny_1983,Kilkenny_1984}.

Optimal control problems are formulated and solved using optimization theories and machine learning platforms. Optimization theories provide rigorous frameworks to solve control problems by finding the optimal control strategies and the underlying Bellman optimality equations or the Hamilton–Jacobi–Bellman (HJB) equations~\cite{Lewis_2012,Bellman1957,AbouheaIJCNN2013,AbouheafCH2014,AbouheafCTT2015}. These solution processes guarantee optimal cost-to-go evaluations. Tracking control mechanism that uses time-varying sliding surfaces is adopted for a two-link manipulator with variable payloads in~\cite{Slotine1983}. It is shown that a~reasonable tracking precision can be obtained using approximate continuous control laws, without~experiencing undesired high frequency signals. An~output tracking mechanism for nonminimum phase flat systems is developed to control the vertical takeoff and landing of an aircraft~\cite{MARTIN1996}. The~underlying state-tracker works well for slightly as well as strongly nonminimum phase systems, unlike the traditional state-based approximate-linearized control schemes. A~state feedback mechanism based on a backstepping control approach is developed for a two-degrees-of-freedom mobile robot. This technique introduced restrictions about the initial tracking errors and the desired velocity of the robot~\cite{JIANG1997}. Observer-based fuzzy controller is employed to solve the tracking control problem of a two-link robotic system~\cite{Tseng2001}. This controller used a convex optimization approach to solve the underlying linear matrix inequality problem to obtain bounded tracking errors~\cite{Tseng2001}. A~state feedback tracking mechanism for underactuated ships is developed in~\cite{Lefeber2003}. The~nonlinear stabilization problem is transformed into equivalent cascaded linear control systems. The~tracking error dynamics are shown to be globally $\cal{K}$- exponentially stable provided that the reference velocity does not decay to zero. An~adaptive neural network scheme is employed to design a cooperative tracking control mechanism where the agents are interacting via a directed communication graph, and~they are tracking the dynamics of a high-order non-autonomous nonlinear system~\cite{ZHANG2012}. The~graph is assumed to be strongly connected and the cooperative control solution is implemented in a distributed fashion. Adaptive backstepping tracking control technique is adopted to control a class of nonlinear systems with arbitrary switching forms in~\cite{ZHAO2015}. It includes an adaptive mechanism to overcome the over-parameterization of the underlying state feedback control laws. A~tracking control strategy is developed for a class of Multi-Input-Multi-Output (MIMO) high-order systems to compensate for the unstructured dynamics in~\cite{Xian2004}. Lyapunov proof with weak assumptions emphasized semi-global asymptotic tracking characteristics of the controller. Fuzzy~adaptive state feedback and observer-based output feedback tracking control architecture is developed for Single-Input-Single-Output (SISO) nonlinear systems in~\cite{Tong2016}. This~structure employed backstepping approach to design the tracking control law for uncertain non-strict feedback~systems.

Machine learning platforms present implementation kits of the derived optimal control mathematical solution frameworks. These use artificial intelligence tools such as Reinforcement Learning (RL) and Neural Networks to solve the Approximate Dynamic Programming problems (ADP)~\cite{Werbos1990,Bertsekas1996,Werbos1974,Werbos1992,Howard_1960,SI2004,Werbos1989}. The~optimization frameworks provide various optimal solution structures which enable solutions of different categories of the approximate dynamic programming problems such as Heuristic Dynamic Programming (HDP), Dual Heuristic Dynamic Programming (DHP), Action~Dependent Heuristic Dynamic Programming (ADHDP), and~Action-Dependent Dual Heuristic Dynamic Programming (ADDHP)~\cite{Abouheapolicy2017,Abouheaf2014}.
These forms in turn are solved using different two-step temporal difference solution structures. ADP approaches provide means to solve the curse of dimensionality in the states and action spaces of the dynamic programming problems.
Reinforcement learning frameworks suggest processes that can implement solutions for the different approximate dynamic programming structures. These are concerned with solving the Hamilton–Jacobi–Bellman equations or Bellman optimality equations of the underlying dynamical structures~\cite{Prokhorov1997,Sutton_1998,Vrancx2008}. 
Reinforcement learning approaches employ dynamic learning environment to decide the best actions associated with the state-combinations to minimize the overall cumulative cost. The~designs of the cost or reward functions reflect the optimization objectives of the problem and play crucial role to find suitable temporal difference solutions~\cite{AbouheafED,Widrow1973,Werbos_1992}.  
This is done using two-step processes, where one solves the temporal difference equation and the other solves for the associated optimal control strategies. Value and policy iteration methods are among the various approaches that are used to implement these steps. The~main differences between the two approaches are related to the sequence of how the solving value functions are evaluated, and the associated control strategies are~updated.

Recently, innovative robust policy and value iteration techniques have been developed for single and multi-agent systems, where the associated computational complexities are alleviated by the adoption of model-free features~\cite{Busoniu2008}.  A~completely distributed model-free policy iteration approach is proposed to solve the graphical games in~\cite{AbouheafCTT2015}. Online policy iteration control solutions are developed for flexible wing aircraft, where approximate dynamic programming forms with gradient structures are used~\cite{AbouheafRV2017,AbouhRob18}. Deep reinforcement learning approaches enable agents to drive optimal policies for high-dimensional environments~\cite{Thanh2018}. Furthermore, they promote multi-agent  collaboration to achieve structured and complex tasks. The~augmented Algebraic Riccati Equation (ARE) for the linear quadratic tracking problem is solved using Q-learning approach in~\cite{Bahare2014}. The~reference trajectory is generated using a linear generator command system. A~neural network scheme based on a reinforcement learning approach is developed for a class of affine (MIMO) nonlinear systems in~\cite{Liu2015}. This approach customized the number of updated parameters irrespective of the complexity of the underlying systems. Integral reinforcement learning scheme is employed to solve the Linear-Quadratic-Regulator (LQR) problem for optimized assistive Human Robot Interaction (HRI) applications in~\cite{Modares2016}. The~LQR scheme optimizes the closed-loop features for a given task to minimize the human efforts without acquiring information about their dynamical models. A~solution framework based on a combined model predictive control and reinforcement learning scheme is developed for robotic applications in~\cite{Zhang2016}. This mechanism uses a guided policy search technique and the model predictive controller generates the training data using the underlying dynamical environment with full state observations. Adaptive control approach based on a model-based structure is adopted to solve the optimal tracking infinite horizon problem for affine systems in~\cite{Kamalapurkar2017}. In~order to effectively explore the dynamical environment, a~concurrent system identification learning scheme is adopted to approximate the underlying Bellman approximation errors. A~reinforcement learning approach based on deep neural networks is used to develop a time-varying control scheme for a formation of unmanned aerial vehicles in~\cite{Conde2017}. The~complexity of the multi-agent structure is tackled by training an individual vehicle and then generalizing the learning outcome of that agent to the formation scheme. Deep Q-Networks are used to develop generic multi-objective reinforcement learning scheme in~\cite{Nguyen2018}. This approach employed single-policy as well as multi-policy structures and it is shown to converge effectively to optimal Pareto solutions. Reinforcement Learning approaches based on deterministic policy gradient, proximal policy optimization, and~trust region policy optimization approaches are proposed to overcome the PID control limitations of the inner attitude control loop of the unmanned aerial vehicles in~\cite{Koch2019}.
The cooperative multi-agent learning systems use the interactions among the agents to accomplish joint tasks in~\cite{Panait2005}. The~complexity of these problems depends on the scalability of the underlying system of agents along with their behavioral objectives.
Action~coordination mechanism based on a distributed constraint optimization approach is developed for multi-agent systems in~\cite{Zhang2013}. It uses an interaction index to trade-off between the beneficial coordination among the agents and the communication cost. This approach enables non-sequenced coupled adaptations of the coordination set and the policy learning processes for the agents.
The mapping of single-agent deep reinforcement learning to multi-agent schemes is complicated due to the underlying scaling dilemma~\cite{Foerster2017}. The~experience replay memory associated with deep Q-learning problems is tackled using a multi-agent sampling mechanism which is based on a variant of importance mechanism in~\cite{Foerster2017}.

The adaptive critics approaches are employed to advise various neural network solutions for optimal control problems. They implement two-step reinforcement learning processes using separate neural network approximation schemes. The~solution for Bellman optimality equation or the Hamilton–Jacobi–Bellman equation is implemented using a feedforward neural structure described by the critic structure. On~the other hand, the~optimal control strategy is approximated using an additional feedforward neural network structure called the actor structure. The~update processes of the actor and critic weights are interactive and coupled in the sense that the~actor weights are tuned when the critic weights are updated following reward/punish assessments of the dynamic learning environment~\cite{Werbos1989,Bertsekas1996,Werbos1992,Sutton_1998,Widrow1973}. The~sequences of the actor and critic weights-updates  follow those advised by the respective value or policy iteration algorithms~\cite{Bertsekas1996,Sutton_1998}. Reinforcement learning solutions are implemented in continuous-time platforms as well as discrete-time platforms, where integral forms of Bellman equations are used~\cite{AbouheafECC14,DragAut}.
These structures are applied to multi-agent systems as well as single-agent systems, where each agent has its own actor-critic structure~\cite{Abouheapolicy2017,Abouheaf2014}.
The adaptive critics are employed to provide neural network solutions for the dual heuristic dynamic programming problems for multi-agent systems~\cite{AbouheafCH2014,AbouheaIJCNN2013}. These structures solve the underlying graphical games in a distributed fashion where the neighbor information is used. Actor-critic solution implementation for an optimal control problem with nonlinear cost function is introduced in~\cite{AbouheafECC14}. The~adaptive critics implementations for feedback control systems are highlighted in~\cite{bahare18}. A~PD scheme is combined with a reinforcement learning mechanism to control the tip-deflection and trajectory-tracking operation of a two-link flexible manipulator in~\cite{Pradhan2012}. The~adopted actor-critic learning structure compensates for the variations in the payload. An~adaptive trajectory-tracking control approach based on actor-critic neural networks is developed for a fully autonomous underwater vehicle in~\cite{Cui2017}. The~nonlinearities in the control input signals are compensated for during the adaptive control~process.

This work contributions are~four-fold: 
	\begin{enumerate}[leftmargin=*,labelsep=5mm]
		\item An online control mechanism is developed to solve the tracking problem in uncertain dynamical environment without acquiring any knowledge about the dynamical models of the underlying systems.
		\item An innovative temporal difference solution is developed using a reformulation of Bellman optimality equation. This form does not require existence of admissible initial policies and it is computationally simple and easy to apply. 
		\item The developed learning approach solves the tracking problem for each dynamical process using separate interactive linear feedback control laws. These optimize the tracking as well as the overall dynamical behavior. 
		\item The outcomes of the proposed architecture can be generalized smoothly for structured dynamical problems. Since, the~learning approach is suitable for discrete-time control environments and it is applicable for complex coupled dynamical problems.    
\end{enumerate}

The paper is structured as follows: Section~\ref{sec:control-problem} is dedicated to the formulation of the optimal tracking control problem  along with the model-free temporal difference solution forms. Model-free adaptive learning processes are developed in Section~\ref{sec:adaptiveprocesses}, and~their real-time adaptive critics or neural network implementations are presented in Section~\ref{sec:NN}. Digital simulation outcomes for an autonomous controller  of a flexible wing aircraft are analyzed in Section~\ref{sec:Autonomous}. The~implications of the developed machine learning processes in practical applications and some future research directions are highlighted in Section~\ref{sec:futurework}. Finally, concluding remarks about the adaptive learning mechanisms are presented in Section~\ref{sec:Conclusion}.

\section{Formulation of the Optimal Tracking Control~Problem}
\label{sec:control-problem}
Optimal tracking control theory is used to lay out the mathematical foundation of various adaptive learning solution frameworks. Thus, many adaptive mechanisms employ complicated control strategies which are difficult to implement in discrete-time solution environments. In~addition, many tracking control schemes are model-dependent, which raises concerns about their performances in unstructured dynamical environments~\cite{Lewis_2012}. This section tackles these challenges by mapping the optimization objectives of underlying tracking problem using machine learning solution~tools.

\subsection{Combined Optimization~Mechanism}

The optimal tracking control problem, in~terms of the operation, can be divided broadly into two main objectives~\cite{Lewis_2012}. The~first is concerned with  asymptotically stabilizing the tracking error dynamics of the system, and~the second optimizes the overall energy during the tracking process. Herein, the~outcomes of the online adaptive learning processes are two linear feedback control laws. The~adaptive approach uses simple linear quadratic utility or cost functions to evaluate the real-time optimal control strategies. The~proposed approach tackles many challenges associated with the traditional tracking problems~\cite{Lewis_2012}. First, it allows an online model-free mechanism to solve the tracking control problem. Second, it allows several flexible tracking control configurations which are adaptable with the complexity of the dynamical systems. Finally, it allows interactive adaptations for both the tracker and optimizer feedback control laws.  

The learning approach does not employ any information about the dynamics of the underlying system. The~selected online measurements can be represented symbolically using the following form
\begin{equation}
X_{k+1}=F(X_k,U_k),
\label{dynmaics}
\end{equation}
where $X \in \mathbb{R}^{n\times 1}$ is a vector of selected measurements (i.e., the~sufficient or observable online measurements), $U \in \mathbb{R}^{m\times1}$ is a vector of control signals, $k$ is a discrete-time index, and~$F$ represents the model that generates the online measurements of the dynamical system which could retain linear or nonlinear~representations.

The tracking segment of the overall tracking control scheme generates the optimal tracking control signal $C_{k\{i\}}\in \mathbb{R} \,\, \forall k$ using a linear feedback control law that depends on the sequence of tracking errors $e_{k\{i\}},e_{{k-1}\{i\}},e_{{k-2}\{i\}},$ 
where each error signal $e_{k\{i\}}$ is associated with the $i^{th}$ state or measured variable of vector $X_{k}$ (i.e., $X_{k\{i\}}$). The~error $e_{k\{i\}}$ is defined by $e_{k\{i\}}=T_{k\{i\}}-X_{k\{i\}},$ where $T_{k\{i\}}$ is the reference signal of the state or measured variable $X_{k\{i\}}$. On~one side, the~number of online tracking control loops is determined by the number of reference variables or states. Each reference signal $T_{k\{i\}}$ has a tracking evaluation loop. In~this development, a~feedback control law that uses combination of three errors (i.e., $e_{k\{i\}},e_{{k-1}\{i\}},e_{{k-2}\{i\}}$) is considered in order to mimic the mechanism of a Proportional-Integral-Derivative (PID) controller in discrete-time where the tracking gains are adapted in real time in an online fashion.  
On the other side, the~form of each scalar tracking control law $C_{k\{i\}}$ can be formulated for any combinations of error samples (i.e., $e_{k\{i\}},e_{{k-1}\{i\}},e_{{k-2}\{i\}},e_{{k-3}\{i\}},\dots, e_{{k-N}\{i\}}$).
Thus, the~proposed tracking structure enables higher-order difference schemes which can be realized smoothly in discrete-time environments. In~order to simplify the tracking notations, $e_{k}$ and $C_{k}$ are used to refer to the tracking error signal $e_{k\{i\}}$ and  tracking control signal $C_{k\{i\}}$ for each individual tracking loop respectively.  Herein, each scalar actuating tracking control signal $C_{k\{i\}}$ simultaneously adjusts all relevant or applicable actuation control signals $U_{k\{j\}},j \in m$.

\begin{figure}[H]
	\centering
	\includegraphics[width=1\textwidth]{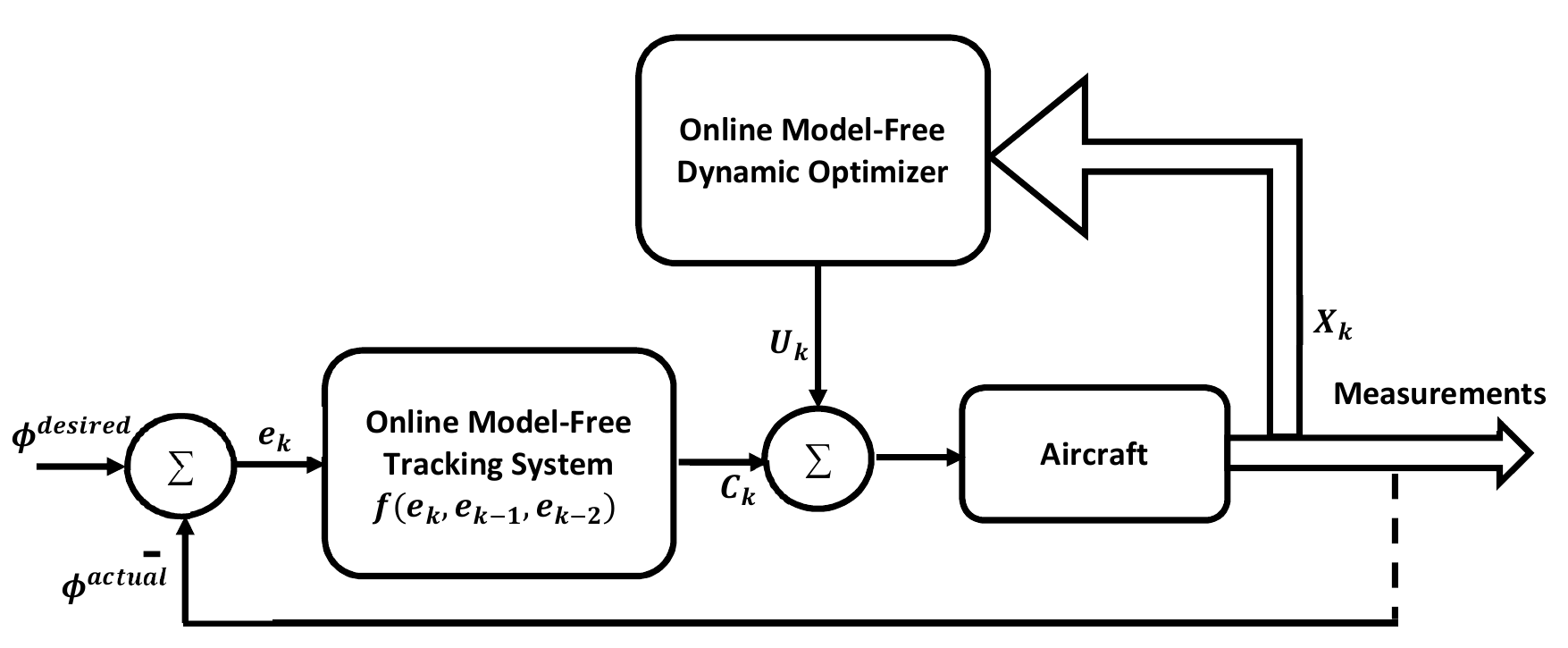}
	\caption{The combined tracking control~mechanism.}
	\label{fig:block}
\end{figure}

The overall layout of the control mechanism (i.e., considering the optimizing and tracking features) is sketched in Figure~\ref{fig:block}, where $\phi^{desired}$ denotes a desired reference signal (i.e., each $T_{k\{i\}}$) and $\phi^{actual}$ refers to the actual measured signal (i.e., each $X_{k\{i\}}$) for each individual tracking~loop. 

The goals of the optimization problem are to find the optimal linear feedback control laws or the optimal control signals $U_k^*$ and $C_k^* \forall k,$ using model-free machine learning schemes. The~underlying objective utility functions are mapped into different temporal difference solution forms. As~indicated above, since linear feedback control laws are used, then linear quadratic utility functions are employed to evaluate the optimality conditions in real time. The~objectives of the optimization problem are detailed as follows:

(1) A measure index of the overall dynamical performance is minimized to calculate the optimal control signal $U_k^*$ such that
$$min_{U_k} \quad O(X_k,U_k)$$
\textls[-15]{with linear quadratic objective cost function $O(X_k,U_k)=\frac{1}{2}\left(X_k^T\,Q\,X_k+U_k^T\,R\,U_k\right),$
where $Q \in \mathbb{R}^{n \times n} > \mathbf{0}$} and $R \in \mathbb{R}^{m \times m} > \mathbf{0}$ are symmetric positive definite~matrices. 

Therefore, the~underlying performance index $J$ is given by
$$J=\sum_{i=k}^{\infty} O\left(X_i,U_i\right).$$

(2) A tracking error index is optimized to evaluate the optimal tracking control signal $C_k^*$ such that
$$min_{C_k} \quad D(E_k,C_k).$$
with an objective cost function $D(E_k,U_k)=\frac{1}{2}\left(E_k^T\,S\,E_k+C_k^T\,M\,C_k\right),$
where $E_k=\left[\begin{array}{ccc}
	{e}_{k}&
	{e}_{k-1}&
	{e}_{k-2}
	\end{array}\right]^T,$ $S \in \mathbb{R}^{3 \times 3} > \mathbf{0}$ is a symmetric positive definite matrix, and~$M \in \mathbb{R} > \mathbf{0}$. The~choice of the tracking error vector $E$ is flexible to the number of the memorized tracking error signals $N_e$ such that \mbox{$e_{k-\ell}, \ell=0, 1,\dots,N_e$}. 

Therefore, the~underlying performance index $P$ is given by
$$P=\sum_{i=k}^{\infty}D\left(E_i,C_i\right).$$

Herein, the~choice of the optimized policy structure $U^*_k$ to be a function of the states $X_k$ is not meant to achieve asymptotic stability in a standalone operation (i.e., all the states $X_k, \forall k$ go to zero). Instead, it is incorporated into the overall control architecture where it can select the minimum energy path during the tracking process. Hence, it creates an asymptotically stable performance around the desired reference trajectory. Later, the~performance of the standalone tracker is contrasted against that of the combined tracking control scheme to highlight this energy exchange minimization~outcome.

\subsection{Optimal Control~Formulation}
Various optimal control formulations of the tracking problem promote multiple temporal difference solution frameworks~\cite{Bellman1957,Lewis_2012}. These use Bellman equations or Hamilton–Jacobi–Bellman structures or even gradient forms of Bellman optimality equations~\cite{AbouheaIJCNN2013,AbouheafCH2014,Abouheaf2014}. The~manner at which the cost or objective function is selected plays a crucial role in forming the underlying temporal difference solution and hence its associated optimal control strategy form. This work provides a generalizable machine learning solution framework, where the optimal control solutions are found by solving the underlying Bellman optimality equations of the dynamical systems. These can be implemented using policy iteration approaches with model-based schemes. However, these processes necessitate having initial admissible policies, which is essential to ensure admissibility of the future policies. This is further faced by computational limitations, for~example, the~reliance of the solutions on least square approaches with possible singularities-related calculation risks. This urged for flexible developments such as online value iteration processes where they do not encounter these~problems.

Value iteration processes based on two temporal difference solution forms are developed to solve the tracking control problem. These are equivalent to solving the underlying Hamilton–Jacobi–Bellman equation of the optimal tracking control problem~\cite{Bahare2014,Lewis_2012}. Regarding the problem under consideration, it is required to have two temporal difference equations: One solves for the optimal control strategies to minimize the tracking efforts, and~the other selects the supporting control signals to minimize the energy exchanges during the tracking process. In~order to do that, two solving value functions related to the main objectives, are proposed such~that

$$\Gamma\left(X_k,U_k\right) =J=\sum_{i=k}^{\infty} O\left(X_i,U_i\right),$$ 
where $\Gamma (\dots)$ is a solving value function that approximates the overall minimized dynamical performance and it is defined by
\begin{eqnarray}
\Gamma\left(X_k,U_k\right) = \displaystyle \frac{1}{2} \, [ {X}_{k} ^\mathrm{T} \,\,\,\,   {U}_{k}^\mathrm{T}]\,\,\,  {H} \,\,\, 
\left[\begin{array}{l}
{X}_{k}\\
{U}_{k}
\end{array}\right], \, {H} =\left[\begin{array} {ll}
{H}_{ {X} {X}}& {H}_{ {X} {U}}\\
{H}_{ {U} {X}}& {H}_{ {U} {U}}
\end{array}\right]. 
\nonumber
\end{eqnarray}

Similarly, the~solving value function that approximates the optimal tracking performance is given~by
$$\Xi\left(E_k,C_k\right) =P=\sum_{i=k}^{\infty}D\left(E_i,C_i\right),$$
where $\Xi\left(E_k,C_k\right) = \displaystyle \frac{1}{2} \, [ {E}_{k} ^\mathrm{T} \,\,\,\,   {C}_{k}^\mathrm{T}]\,\,\,  {\Pi} \,\,\, 
\left[\begin{array}{l}
{E}_{k}\\
{C}_{k}
\end{array}\right], \, {H} =\left[\begin{array} {ll}
{\Pi}_{ {E} {E}}& {\Pi}_{ {E} {C}}\\
{\Pi}_{ {C} {E}}& {\Pi}_{ {C} {C}}
\end{array}\right].$

These performance indices yield the following Bellman or temporal difference equations
\begin{equation}
\Gamma\left(X_k,U_k\right) =\frac{1}{2}\left(X_k^T\,Q\,X_k+U_k^T\,R\,U_k\right)+\Gamma\left(X_{k+1},U_{k+1}\right),
\label{Bell1}
\end{equation}
 and
\begin{equation}
\Xi\left(E_k,C_k\right) =\frac{1}{2}\left(E_k^T\,S\,E_k+C_k^T\,M\,C_k\right)+\Xi\left(E_{k+1},C_{k+1}\right).
\label{Bell2}
\end{equation}
where the optimal control strategies associated with both Bellman equations are calculated as follows
$$U^*_k=argmin_{U_k} \Gamma\left(X_k,U_k\right) \quad \rightarrow \quad H_{UX}\, X_k+H_{UU}\, U^*_k=0.$$

Therefore, the~optimal policy for the overall optimized performance is given by
\begin{equation}
U^*_k=- \, H^{-1}_{UU} \, H_{UX}\, X_k.
\label{op1}
\end{equation}

In a similar fashion, the~optimal tracking control strategy is calculated using 
$$C^*_k=argmin_{C_k} \Xi\left(E_k,C_k\right) \quad \rightarrow \quad \Pi_{CE}\, E_k+\Pi_{CC}\, U^*_k=0.$$

Therefore, the~optimal policy for the optimized tracking performance is given by
\begin{equation}
C^*_k=- \, \Pi^{-1}_{CC} \, \Pi_{CE}\, E_k.
\label{op2}
\end{equation}

Using the optimal policies (\ref{op1}) and (\ref{op2}) into Bellman Equations~(\ref{Bell1}) and (\ref{Bell2}) respectively yields the following Bellman optimality equations or temporal difference equations
\begin{equation}
\Gamma^*\left(X_k,U^*_k\right) =\frac{1}{2}\left(X_k^T\,Q\,X_k+U_k^{*T}\,R\,U^*_k\right)+\Gamma^*\left(X_{k+1},U^*_{k+1}\right),
\label{OBell1}
\end{equation}
and
\begin{equation}
\Xi^*\left(E_k,C^*_k\right) =\frac{1}{2}\left(E_k^T\,S\,E_k+C_k^{*T}\,M\,C^*_k\right)+\Xi^*\left(E_{k+1},C^*_{k+1}\right).
\label{OBell2}
\end{equation}
where $\Gamma^*(\dots)$ and $\Xi^*(\dots)$ are the optimal solutions for the above Bellman optimality~equations

Solving Bellman optimality Equations~(\ref{OBell1}) or (\ref{OBell2}) is equivalent to solving the underlying Hamilton–Jacobi–Bellman equations of the optimal tracking control~problem.  

\begin{Remark}
Model-free value iteration processes employ temporal difference solution forms that arise directly from Bellman optimality Equations~(\ref{OBell1}) or (\ref{OBell2}), in~order to solve the proposed optimal tracking control problem. This learning platform shows how to enable Action-Dependent Heuristic Dynamic Programming (ADHDP) solution, a~class of approximate dynamic programming that employs a solving value function that is dependent on a state-action structure, in~order to solve the optimal tracking problem in an online fashion~\cite{Sutton_1998,Landelius1996}. 
\end{Remark}

\section{Online Model-Free Adaptive Learning~Processes}
\label{sec:adaptiveprocesses}
Bellman optimality Equations~(\ref{OBell1}) and (\ref{OBell2}) are used to develop online value iteration processes. Herein, two adaptive learning algorithms are developed using these optimality equations. They share the ability to produce control strategies while they learn the dynamic environment in real time and the strategies do not depend on the dynamical model of the system under~consideration.

\subsection{Direct Value Iteration~Process}

The first model-free value iteration algorithm uses direct forms of (\ref{OBell1}) and (\ref{OBell2}) as follows:

\begin{algorithm}[H]
	\caption{Model-free direct value iteration~process.}\label{alg:alg1}
	\begin{algorithmic}
		\State
		\begin{enumerate}[leftmargin=*,labelsep=5mm]
			\item Initialize $\Gamma^0(X_0, U_0),\,\Xi^0(E_0, C_0),U_{0}^0$ and $C_{0}^0$. \vspace{10pt}
			\item {Update the solving value functions $\Gamma (\dots)$ and $\Xi(\dots)$ using}
\begin{eqnarray}
\Gamma^{r+1}\left(X_k,U_k\right)&=&O^r\left(X_k,U_k\right)+\Gamma^r\left(X_{k+1},U_{k+1}\right),
\nonumber
\\
\Xi^{r+1}\left(E_k,C_k\right)&=&D^r\left(E_k,C_k\right)+\Xi^r\left(E_{k+1},C_{k+1}\right),
\label{val_bell1}
\end{eqnarray}
where $r$ is an evaluation index.
\vspace{10pt}
\item Extract the optimal strategies
\begin{eqnarray}
			U^{r+1}_k&=&- \, \left[H^{-1}_{UU} \, H_{UX}\right]^{r+1}\, X_k,
			\nonumber
\\			C^{r+1}_k&=&- \, \left[\Pi^{-1}_{CC} \, \Pi_{CE}\right]^{r+1}\, E_k.
\label{val-pol1}
			\end{eqnarray}
			\item Terminate the updates of the solving value functions when $\Vert H^{r+1}(..) - H^{r}(..)\Vert \le \varepsilon$ and $\Vert \Pi^{r+1}(..) - \Pi^{r}(..)\Vert \le \varepsilon$, $\varepsilon$ is an error threshold.
		\end{enumerate}
	\end{algorithmic}
\end{algorithm}

\subsection{Modified Value Iteration~Process}

Another adaptive learning algorithm based on an indirect value iteration process is proposed. This algorithm reformulates or modifies the way Bellman optimality equations are solved as follows;
\begin{equation}
\Gamma^*\left(X_k,U^*_k\right)-\Gamma^*\left(X_{k+1},U^*_{k+1}\right) =\frac{1}{2}\left(X_k^T\,Q\,X_k+U_k^{*T}\,R\,U^*_k\right),
\label{OBell3}
\end{equation}
and
\begin{equation}
\Xi^*\left(E_k,C^*_k\right)-\Xi^*\left(E_{k+1},C^*_{k+1}\right) =\frac{1}{2}\left(E_k^T\,S\,E_k+C_k^{*T}\,M\,C^*_k\right).
\label{OBell4}
\end{equation}

Therefore, a~modified value iteration process based on these reformulations is structured as~follows
\begin{algorithm}[H]
	\caption{Modified model-free value iteration~process.}\label{alg:alg2}
	\begin{algorithmic}
		\State
		\begin{enumerate}[leftmargin=*,labelsep=5mm]
			\item Initialize $\Gamma^0(X_0, U_0),\,\Xi^0(E_0, C_0),U_{0}^0$ and $C_{0}^0$. \vspace{10pt}
			\item {Update the solving value functions $\Gamma (\dots)$ and $\Xi(\dots)$ using}
\begin{eqnarray}
			\Gamma^{r+1}\left(X_k,U_k\right)-\Gamma^{r+1}\left(X_{k+1},U_{k+1}\right)&=&O^r\left(X_k,U_k\right),
			\nonumber
\\
			\Xi^{r+1}\left(E_k,C_k\right)-\Xi^{r+1}\left(E_{k+1},C_{k+1}\right)&=&D^r\left(E_k,C_k\right).
\label{val_bell2}
			\end{eqnarray}
			\item Extract the optimal strategies
\begin{eqnarray}
			U^{r+1}_k&=&- \, \left[H^{-1}_{UU} \, H_{UX}\right]^{r+1}\, X_k,
			\nonumber
\\
			C^{r+1}_k&=&- \, \left[\Pi^{-1}_{CC} \, \Pi_{CE}\right]^{r+1}\, E_k.
\label{val_pol2}
			\end{eqnarray}
			\item Terminate the updates of the solving value functions when $\Vert H^{r+1}(..) - H^{r}(..)\Vert \le \varepsilon$ and $\Vert \Pi^{r+1}(..) - \Pi^{r}(..)\Vert \le \varepsilon$.
		\end{enumerate}
	\end{algorithmic}
\end{algorithm}
This value iteration process solves Bellman optimality equation in a way that does  not require initial stabilizing policies and, unlike the policy iteration mechanisms, this solution framework does not imply any computational difficulties related to the evaluations of $\Gamma(\dots)$ and $\Xi(\dots)$ at the different evaluation~steps.

The proposed value iteration processes optimize the overall dynamical performance towards the tracking objectives. This means that the two optimization objectives are interacting and coupled along the variables of interest. This is done in real time without acquiring any information about the dynamics of the underlying~system.

\subsection{Comparison to a Standard Policy Iteration~Process}
The value iteration process, as~explained earlier, employs two steps, one is concerned with evaluating the optimal value function (i.e., solving Bellman optimality Equations~(\ref{OBell1}) or (\ref{OBell2})) and the second extracts the optimal policy given this value function (i.e., (\ref{op1}) or (\ref{op2})). On~the other hand, the~policy iteration mechanism starts with a policy evaluation step that solves for a value function that is relevant to an attempted policy using Bellman equation (i.e., (\ref{Bell1}) or (\ref{Bell2})) and this is followed by a policy improvement step that results in a strictly better policy compared to the preceding policy unless it is optimal~\cite{Sutton_1998,lewis2009,DragAut}.

To formulate a policy iteration process for the optimization problem in hand (\mbox{i.e., the overall} energy and tracking error minimization), the~control signals $U^{H}$ and $C^{\Pi}$ are evaluated using the linear policies $-\left[H^{-1}_{UU} \, H_{UX}\right] \, X$ and $-\left[\Pi^{-1}_{CC} \, \Pi_{CE}\right] \, E$, respectively, where the policy iteration process uses (\ref{Bell1}) and (\ref{Bell2}) repeatedly in order to perform a single-policy evaluation step, such that
\begin{eqnarray}
\Gamma^{j}\left(X_k,U^{H}_k\right)-\Gamma^{j}\left(X_{k+1},U^{H}_{k+1}\right)&=&O\left(X_k,U^{H}_k\right),
\nonumber
\\
\Xi^{h}\left(E_k,C^{\Pi}_k\right)-\Xi^{h}\left(E_{k+1},C^{\Pi}_{k+1}\right)&=&D\left(E_k,C^{\Pi}_k\right),
\nonumber
\end{eqnarray}
where the symbols $j$ and $h$ refer to the calculation-instances leading to a policy evaluation step for each dynamical~operation.

In other words, the~solving value function $\Gamma(\dots)$ is updated after collecting several necessary samples $\nu$  $\left( \text{i.e.,} \, \tilde Z_X^{j=1}(X_{k,k+1},U^H_{k,k+1}),\right.$ $\tilde Z_X^{j=2}(X_{k+1,k+2},U^H_{k+1,k+2}), $ $ \dots,$ $\left. \tilde Z_X^{j=\nu}(X_{k+\nu-1,k+\nu},U^H_{k+\nu-1,k+\nu})\right),$ where $\nu=(n+m)\times(n+m+1)/2$ designates the number of entries of the upper/lower triangle block of matrix $H \in \mathbb{R}^{(n+m)\times(n+m)}$ and $\tilde Z_X$ is a vector associated with the vector transformation of the upper/lower triangle block of the symmetric matrix $H$~\cite{lewis2009,DragAut}. This act lasts for at least a real-time interval of $k$ to $k+\nu$ to collect sufficient information to fulfill the policy evaluation step~\cite{lewis2009,DragAut}. 
Similarly, the~solving value function $\Xi\left(\dots\right)$ is updated at the end of each online interval $k$ to $k+10$, where $10$ samples (10 refers to the number of entries of the upper/lower triangle block of matrix \mbox{$\Pi \in \mathbb{R}^{4\times4}$}) are repeatedly collected in order to evaluate the taken tracking policy $\left( \text{i.e.,}\, \tilde Z_E^{h=1}(E_{k,k+1},C^\Pi_{k,k+1}),\right.$ $\tilde Z_E^{h=2}(E_{k+1,k+2},C^\Pi_{k+1,k+2}),$ $  \dots, $ $\left. \tilde Z_E^{h=10}(E_{k+9,k+10},C^\Pi_{k+9,k+10})\right)$, where~the vector $\tilde Z^h_E$ is structured in a similar manner as $\tilde Z_X$. The~approach taken to construct vector $\tilde Z_{X}$ or $\tilde Z_{E}$ is detailed in~\cite{lewis2009,DragAut}. The~policy iteration solution results in a decreasing sequence of the solving value functions which is lower-bounded by~zero.

The policy iteration process requires the existence of an initial admissible policy and could encounter mathematical risks when evaluating the underlying policies~\cite{lewis2009,DragAut}.  
On the other hand, Algorithms~\ref{alg:alg1}~and~\ref{alg:alg2} do not impose initial admissible policies and the optimal value functions $\Gamma(\dots)$ and $\Xi(\dots)$ are updated simultaneously at each real-time instance $r=k$, as~explained by~(\ref{val_bell1})~and~(\ref{val_bell2}). The~value iteration process retains simpler and flexible adaptation mechanism compared with the above policy iteration formulation, where the policy evaluation steps could exist at uncorrelated~time-instances.

\subsection{Convergence and Stability Results of the Adaptive Learning~Mechanism}

The convergence analysis and stability characteristics of the value iteration processes, based on action-dependent heuristic dynamic programming solution, are introduced for single and multi-agent systems and for continuous as well as discrete-time environments~\cite{CDC18,ICRA19,Landelius1996,AbouheafCH2014,Abouheaf2014}.  The~adaptive learning value iteration processes result in non-decreasing sequences such that
\begin{eqnarray}
0 < \dots \le \Gamma^{0}\le\Gamma^{1}\le \Gamma^{2}\le \dots \le \Gamma^{r} \dots \le \Gamma^{*},
\nonumber
\\
0 < \dots \le \Xi^{0}\le\Xi^{1}\le \Xi^{2}\le \dots \le \Xi^{r} \dots \le \Xi^{*},
\nonumber
\end{eqnarray}
where $\Gamma^*(\dots)$ and $\Xi^*(\dots)$ are the upper bounded optimal solutions for Bellman optimality~equations.

The sequences of the resultant control strategies $U_k^r, \forall k,r$ and $C_k^r,\forall k,r$ are stabilizing and hence admissible sequences. In~a similar fashion, the~following inequalities hold
\begin{eqnarray}
\Gamma^{r}(X_k,U_k)-\Gamma^{r}(X_{k+1},U_{k+1})\le\Gamma^{r+1}(X_k,U_k)-\Gamma^{r+1}(X_{k+1},U_{k+1}), \nonumber \\
\Xi^{r}(X_k,U_k)-\Xi^{r}(X_{k+1},U_{k+1})\le\Xi^{r+1}(X_k,U_k)-\Xi^{r+1}(X_{k+1},U_{k+1}). \nonumber 
\end{eqnarray}

The above inequalities are also bounded above using the same concepts adopted in~\cite{CDC18,ICRA19,Landelius1996,AbouheafCH2014,Abouheaf2014}. The~simulation results highlight the evolution of the solving value functions using Algorithms~\ref{alg:alg1}~and~\ref{alg:alg2} in real time. Furthermore, they will judge the importance of Algorithm~~\ref{alg:alg2} in terms of the convergence speed and optimality of the solving value~functions.

\section{Neural Network~Implementations}
\label{sec:NN}
\begin{figure}[H]
	\centering
	\includegraphics[width=0.85\textwidth]{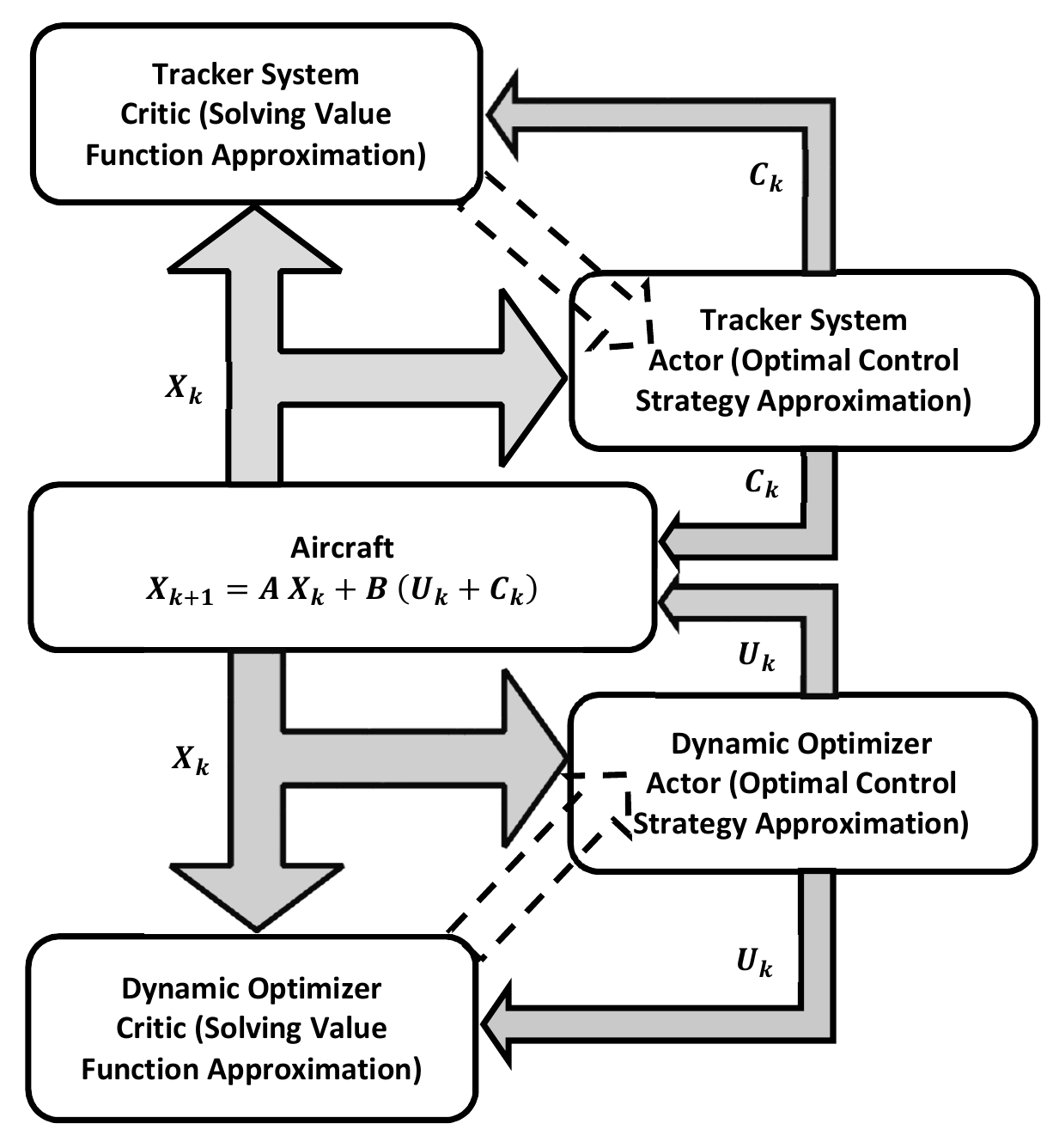}
	\caption{The adaptive critics~scheme.}
	\label{fig:actcrt}
\end{figure}

Adaptive critics are employed to implement the proposed adaptive learning solutions in real time. Each algorithm involves two steps. The~first is concerned with solving a Bellman optimality equation, and~the other approximates the optimal control strategy. Each step is implemented using a neural network approximation structure. The~solving value function $\Gamma (\dots)$ or $\Xi(\dots)$ is approximated using a critic structure, while the associated optimal control policy is approximated using an actor structure. These represent coupled tuning processes with different objectives. The~solving algorithms employ update processes to tune the critic weights, where they have different forms of the temporal difference equations. However, the~way the actor is approximated for both adaptive algorithms is achieved in the same fashion. A~full adaptive critics solution structure for the tracking control problem is shown in Figure~\ref{fig:actcrt}.

\subsection{Neural Network Implementation of Algorithm~1}
The actor-critic adaptations for Algorithm \ref{alg:alg1} are done in real time using separate neural network structures as~follows.

The solving value functions $\Gamma (\dots)$ and $\Xi(\dots)$ are approximated using the neural network structures 
\begin{equation*}
\hat \Gamma (.|\Upsilon_c ) =  \frac{1}{2} [X_{k} ^T \,\,\,\,  \hat U_{k}^T]\,\,\, \Upsilon_c^\mathrm{T} \,\,\, 
\left[\begin{array}{l}
X_{k}\\
\hat U_{k}
\end{array}\right] \quad \text{and} \quad
\hat \Xi (.|\Omega_c ) =  \frac{1}{2} [E_{k} ^T \,\,\,\,  \hat C_{k}^T]\,\,\, \Omega_c^\mathrm{T} \,\,\, 
\left[\begin{array}{l}
E_{k}\\
\hat C_{k}
\end{array}\right],
\end{equation*}
where
$\Upsilon^T_c =  \left[\begin{array} {cc}
\Upsilon^T_{c XX} & \Upsilon^T_{c X\hat U}\\
& \\
\Upsilon^T_{c \hat UX} & \Upsilon^T_{c \hat U\hat U}
\end{array}\right] \in \mathbb{R}^{(n+m) \times (n+m)}$ and $\Omega^T_c =  \left[\begin{array} {cc}
\Omega^T_{c EE} & \Omega^T_{c E\hat C}\\
& \\
\Omega^T_{c \hat CE} & \Omega^T_{c \hat C\hat C}
\end{array}\right] \in \mathbb{R}^{4 \times 4}$
are the critic approximation weights~matrices.

The optimal strategies $U^*$ and $C^*$ are approximated as
\begin{equation*}
\hat U_k= \Upsilon_a \, X_k \quad \text{and} \quad \hat C_k= \Omega_a \, E_k,
\end{equation*}
where $\Upsilon^T_a \in \mathbb{R}^{m \times 1}$
and $\Omega^T_a \in \mathbb{R}^{3 \times 1}$ are the approximation weights of the~actors.

The tuning processes are interactive, and~the weights of each structure are updated using a gradient descent approach. Therefore, the~update laws for the critic weights for this algorithm are calculated as
\begin{eqnarray}
\Upsilon_c^{(r+1)T}&=&\Upsilon_c^{r\,T}-\alpha_c\,\left(\hat \Gamma(.|\Upsilon_c^{r\,T}) - \hat \Gamma^{target}(.|\Upsilon_c^{r\,T})\right) Z_X\, Z_X^T, \nonumber \\ 
\Omega_c^{(r+1)T}&=&\Omega_c^{r\,T}-\alpha_c\,\left(\hat \Xi(.|\Omega_c^{r\,T}) - \hat \Xi^{target}(.|\Omega_c^{r\,T})\right) Z_E\, Z_E^T,
\label{critic1}
\end{eqnarray}
where $0<\alpha_c <1$ is a critic learning rate, $Z_X=\left[\begin{array}{l}
X_{k}\\
\hat U^r_{k}
\end{array}\right]$, 
$Z_E=\left[\begin{array}{l}
E_{k}\\
\hat C^r_{k}
\end{array}\right],$ and the target values of the approximations $\Gamma^{target}(\dots)$ and $\Xi^{target}(\dots)$ are given~by

\begin{eqnarray*}
\Gamma^{target}&=&O\left(X_k,\hat U^r_k\right)+\Gamma^r\left(X_{k+1},\hat U^r_{k+1}\right), \\
\Xi^{target}&=&D\left(E_k,\hat C^r_k\right)+\Pi^r\left(E_{k+1},\hat C^r_{k+1}\right).
\end{eqnarray*}

In a similar fashion, the~approximation weights of the optimal control strategies are updated using the rules

\begin{eqnarray}
\Upsilon_a^{(r+1)T}&=&\Upsilon_a^{r\,T}-\alpha_a\,\left(\hat U - \hat U^{target}\right)^r X_k^T, \nonumber \\ 
\Omega_a^{(r+1)T}&=&\Omega_a^{r\,T}-\alpha_a\,\left(\hat C_k - \hat C_k^{target}\right)^r E_k^T,
\label{actor1}
\end{eqnarray}
where $0<\alpha_a<1$ defines the actor learning rate and the target values of the optimal policy approximations $\hat U_k$ and $\hat C_k$  are given by
\begin{eqnarray}
\hat U^{target}&=&- \, \left[\Upsilon^{-1}_{c\hat U \hat U} \, \Upsilon_{c\hat UX}\right]^r\, X_k, \nonumber \\ 
\hat C^{target}&=&- \, \left[\Omega^{-1}_{c\hat C \hat C} \, \Omega_{c\hat CE}\right]^r\, E_k. \nonumber 
\end{eqnarray}

Consequently, the~critic and actor update laws are given by  (\ref{critic1}) and (\ref{actor1}) respectively, where they form the implementation platforms of the solution steps (\ref{val_bell1}) and (\ref{val-pol1}) in Algorithm \ref{alg:alg1}. 
 
 \begin{Remark}
 The gradient descent approach employs actor-critic learning rates which take positive values less than 1. In~the proposed development the actor-critic learning rates are tied to the sampling time used to generate the online measurements in the discrete-time environment. This is done to achieve smooth tuning for the actor-critic weights relative to the changes in the dynamics of the system.
The gradient decent approaches do not have affirmative convergence criteria. However, as~will be shown below, the~simulation cases emphasize the usefulness of this approach even when a challenging dynamical environment is considered, where~one of the challenging scenarios considers random actor-critic learning rates at each evaluation step in the real time~processes.
\end{Remark}
\subsection{Neural Network Implementation of Algorithm~2}

The following development introduces the neural network implementations of the solution given by the modified value iteration solution presented by Algorithm \ref{alg:alg2}. 

The solving value function approximations $\tilde \Gamma (.|\Delta_c)$ and $\tilde \Xi (.|\Lambda_c)$ are given~by

\begin{equation*}
\tilde \Gamma (.|\Delta_c ) =  \frac{1}{2} [X_{k} ^T \,\,\,\,  \tilde U_{k}^T]\,\,\, \Delta_c^\mathrm{T} \,\,\, 
\left[\begin{array}{l}
X_{k}\\
\tilde U_{k}
\end{array}\right] \quad \text{and} \quad
\tilde \Xi (.|\Lambda_c ) =  \frac{1}{2} [E_{k} ^T \,\,\,\,  \tilde C_{k}^T]\,\,\, \Lambda_c^\mathrm{T} \,\,\, 
\left[\begin{array}{l}
E_{k}\\
\tilde C_{k}
\end{array}\right],
\end{equation*}
where
$\Delta^T_c =  \left[\begin{array} {cc}
\Delta^T_{c XX} & \Delta^T_{c X\tilde U}\\
& \\
\Delta^T_{c \tilde UX} & \Delta^T_{c \tilde U\tilde U}
\end{array}\right] \in \mathbb{R}^{(n+m) \times (n+m)}$ and $\Lambda^T_c =  \left[\begin{array} {cc}
\Lambda^T_{c EE} & \Lambda^T_{c E\tilde C}\\
& \\
\Lambda^T_{c \tilde CE} & \Lambda^T_{c \tilde C\tilde C}
\end{array}\right] \in \mathbb{R}^{4 \times 4}$
are the critic approximation weights~matrices.

The approximations of the optimal control strategies $U^*$ and $C^*$ follow 
\begin{equation*}
\tilde U_k= \Delta_a \, X_k \quad \text{and} \quad \tilde C_k= \Lambda_a \, E_k,
\end{equation*}
where $\Delta^T_a \in \mathbb{R}^{m \times 1}$
and $\Lambda^T_a \in \mathbb{R}^{3 \times 1}$ are the approximation weights of the actor neural~network.

The tuning of the critic weights for both optimization loops follows
\begin{eqnarray}
\bar\Delta_c^{(r+1)T}&=&\bar \Delta_c^{r\,T}-\eta_c\,\left(\tilde \Gamma(.|\Delta_c^{r\,T}) - \tilde \Gamma^{target}(.|\Delta_c^{r\,T})\right) \,\bar  Z_X^T, \nonumber \\ 
\bar\Lambda_c^{(r+1)T}&=&\bar\Lambda_c^{r\,T}-\eta_c\,\left(\tilde \Xi(.|\Lambda_c^{r\,T}) - \tilde \Xi^{target}(.|\Lambda_c^{r\,T})\right) \, \bar Z_E^T,
\label{critic2}
\end{eqnarray}
where $0<\eta_c<1$ is a critic learning rate, $\bar \Delta_c$ and $\bar \Lambda_c$ are vector transformations of the upper triangle  section of the symmetric solution matrices  $\Delta_c$ and $\Lambda_c$ respectively,  $\tilde Z_X$ and $\tilde Z_E$ are the respective vector-to-vector transformations of $\tau_X^r$ and $\tau_E^r$ with 
$\tau_X^r=\left[\begin{array}{l}
X_{k}\\
\tilde U^r_{k}
\end{array}\right]-\left[\begin{array}{l}
X_{k+1}\\
\tilde U^r_{k+1}
\end{array}\right]$ and 
\mbox{$\tau_E^r=\left[\begin{array}{l}
E_{k}\\
\tilde C^r_{k}
\end{array}\right]-\left[\begin{array}{l}
E_{k+1}\\
\tilde C^r_{k+1}
\end{array}\right]$.} 

The target values $\tilde\Gamma^{target} (\dots)$ and $\tilde\Xi^{target} (\dots)$ are calculated by

\begin{eqnarray*}
\tilde 	\Gamma^{target} &=& O\left(X_k,\hat U^r_k\right), \\
\tilde \Xi^{target} &=& D\left(E_k,\hat C^r_k\right).
\end{eqnarray*}

The update of the actor weights for this solution algorithm follows a similar structure as of Algorithm \ref{alg:alg1} such that
\begin{eqnarray}
\Delta_a^{(r+1)T}&=&\Delta_a^{r\,T}-\eta_a\,\left(\tilde U - \tilde U^{target}\right)^r X_k^T, \nonumber \\ 
\Lambda_a^{(r+1)T}&=&\Lambda_a^{r\,T}-\eta_a\,\left(\tilde C_k - \tilde C_k^{target}\right)^r E_k^T,
\label{actor2}
\end{eqnarray}
where $0<\eta_a<1$ is an actor learning rate, and~the target values $\tilde U^{target}(\dots)$ and $\tilde C_k^{target}(\dots)$ are given~by
\begin{eqnarray}
\tilde U^{target}&=&- \, \left[\Delta^{-1}_{c\tilde U \tilde U} \, \Delta_{c\tilde UX}\right]^r\, X_k, \nonumber \\ 
\tilde C^{target}&=&- \, \left[\Lambda^{-1}_{c\tilde C \tilde C} \, \Lambda_{c\tilde CE}\right]^r\, E_k. \nonumber 
\end{eqnarray}

\section{Autonomous Flexible Wing Aircraft~Controller}
\label{sec:Autonomous}
The proposed online adaptive learning approaches are employed to design an autonomous trajectory-tracking controller for a flexible wing aircraft.
The flexible wing aircraft functions as a two-body system (i.e., the pilot/fuselage and wing systems)~\cite{Blake_1991,Cook2006,Cook_Kilkenny_1986,DE_MATTEIS_1990,De_Matteis_1991}.
Unlike fixed wing systems, the~flexible wing aircraft do not have exact aerodynamic models, due to the deformations in the wings which are continuously occurring~\cite{Cook2006,Cook_2013,Ochi_2017}. Aerodynamic modeling attempts rely on semi-experimental results with no exact models, which complicated the autonomous control task and made it very challenging~\cite{Cook2006}. Recently, these aircraft have captured increasing attention to join the unmanned aerial vehicles family due to their low-cost operation features, uncomplicated design, and~simple fabrication process~\cite{AbouhRob18}.
The maneuvers are achieved by changing the relative centers of gravity between the pilot and wing systems. In~order to change the orientation of the wing with respect to the pilot/fuselage system, the~control bar of the aircraft takes different pitch-roll commands to achieve the desired trajectory. The~pitch/roll maneuvers are achieved by applying directional forces on the control bar of the flexible wing system in order to create or alter the desired orientation of the wing with respect to the pilot/fuselage system~\cite{Ochi_2015,Ochi_2017}.

\begin{figure}[H]
	\centering
	\includegraphics[width=.35\textwidth]{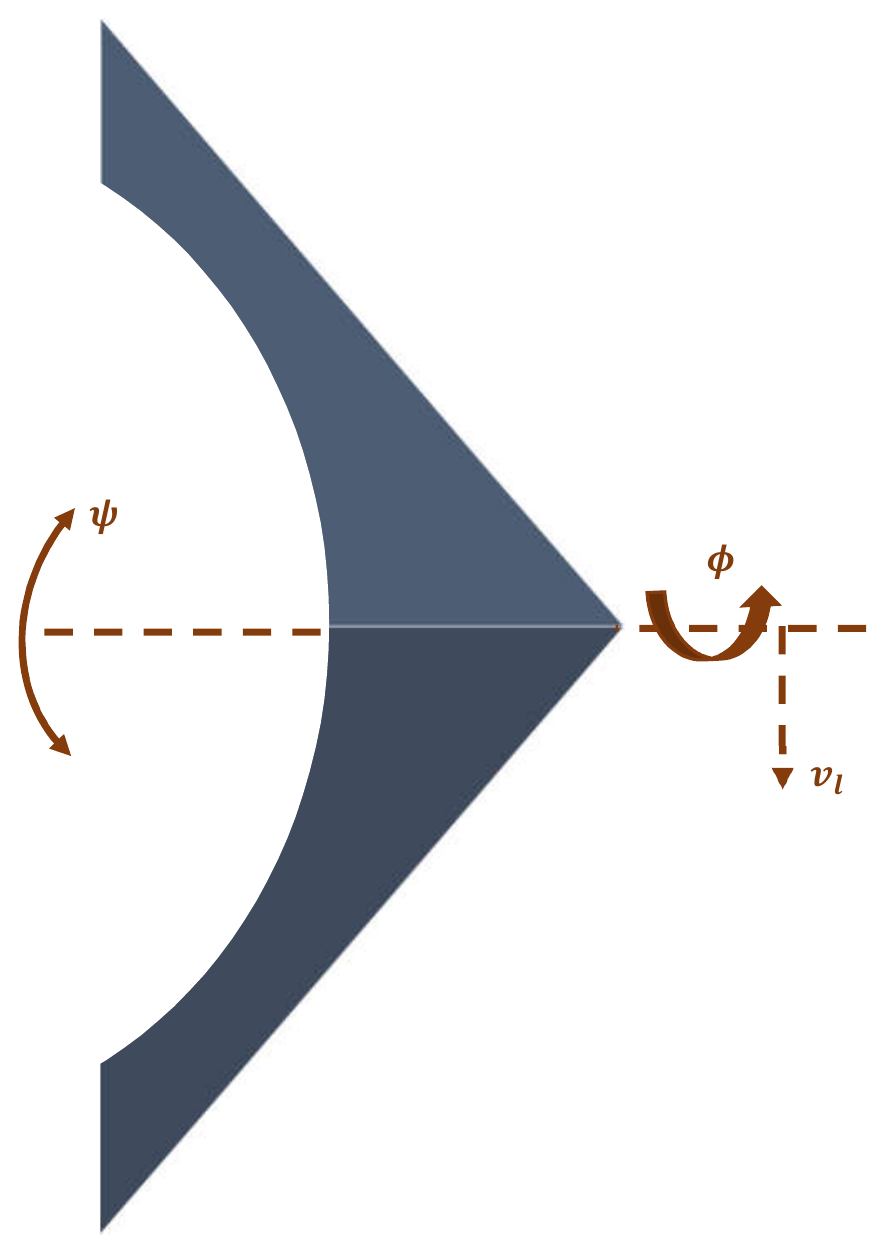}
	\caption{Aircraft wing motion~frame.}
	\label{fig:aircraft}
\end{figure}

The objective of the autonomous aircraft controller design is to use the proposed online adaptive learning structures in order to achieve the roll-trajectory-tracking objectives, and~to minimize energy paths (the dynamics of the aircraft) during the tracking process. The~energy minimization is crucial for the economics of flying systems that share the same optimization objectives. The~motions of the flexible wing aircraft are decoupled into longitudinal and lateral frames~\cite{Cook2006,Cook_2013}. The~lateral motion frame is hard to control compared to the inherited stability in the pitch motion frame. A~lateral motion frame of a flexible wing aircraft is shown in Figure~\ref{fig:aircraft}.

	\subsection{Assessment Criteria for the Adaptive Learning~Algorithms}
	The effectiveness of the proposed online model-free adaptive learning mechanisms is assessed based on the following~criteria:  
	\begin{itemize}[leftmargin=*,labelsep=5.8mm]
		\item The convergence of the online adaptation processes (i.e., tuning of the actor and critic weights achieved using Algorithms \ref{alg:alg1} and \ref{alg:alg2}). Consequently, the~resulting trajectory-tracking error~characteristics.
		\item The performance of the standalone tracking system versus the overall or combined tracking control scheme.  
		\item The stability results of the online combined tracking control scheme (i.e., the aircraft is required to achieve the trajectory-tracking objective in addition to minimizing the energy exchanges during the tracking process).
		\item The benefits of the attempted adaptive learning approaches on improving the closed-loop time-characteristics of the aircraft during the navigation process. 
	\end{itemize}
	
Additionally, the~simulation cases are designed to show how broadly Algorithm \ref{alg:alg2} (i.e.,~the~newly modified Bellman temporal difference framework) will perform against Algorithm \ref{alg:alg1}.

\unskip

\subsection{Generation of the Online~Measurements}
To apply the proposed adaptive approaches on the lateral motion frame, a~simulation environment is needed to generate the online measurements. The~different control methodologies do not use all the available measurements to control the aircraft~\cite{Cook2006,Ochi_2017}. Thus, the~proposed approach is flexible to the selection of the key measurements. Hence, a~lateral aerodynamic model at a trim speed, based on a semi-experimental study, is employed to generate the measurements as follows~\cite{Cook2006}
\begin{eqnarray}
X_{k+1}=A \, X_k+B \,U_{Tk},
\nonumber
\end{eqnarray}
where the lateral state vector of the wing system is given by $X=\begin{bmatrix*}[c]
v_{l} & \dot \phi & \dot \psi & \phi & \psi
\end{bmatrix*}^T$ and $U_T$ is the lateral control signal applied to the control~bar.

The control signal $U_T$ is the overall combined control strategy decided by the tracker system and the optimizer system (i.e., $U_{Tk}=U_{k}+C_{k}$). In~this example, the~banking control signal aggregates dynamically the scalar signals $U_k \, \in \, \mathbb{R}$ and $C_k \, \in \, \mathbb{R}$ in real time in order to get an equivalent control signal $U_{Tk}$ that is applied to the control bar in order to optimize the motion following a trajectory-tracking command. The~optimizer will decide the state feedback control policy $U_{k}=f(X_k)$ using the measurements $X_k$, where the linear state feedback optimizer control gains $\Omega_a, \Lambda_a \in \mathbb{R}^{1 \times 5}$ are decided by the proposed adaptive learning algorithms. Similarly, the~tracking system will decide the linear tracking feedback control policy $C_k$ based on the error signals $(e_k, e_{k-1},e_{k-2})$, where $e_k=\phi^{desired}_k-\phi^{actual}_k, \forall k$. The~linear feedback tracking control gains $\Upsilon_a, \Delta_a \in \mathbb{R}^{1 \times 3}$ are adapted in real time using the online reinforcement learning~algorithms.  

Noticeably, the~proposed online learning solutions do not employ any information about the dynamics (i.e., drift dynamics $A$ and control input matrix $B$), where they function like black-box mechanisms. Moreover, the~control objectives are implemented in an online fashion, where only real-time measurements are considered. In~other words, the~control mechanism for the roll maneuver generates the real-time control strategy for the roll motion frame regardless what is occurring in the pitch direction and vice~versa.

\subsection{Simulation~Environment}

As described earlier, a~state space model captured at a trim flight condition is used to generate online measurements~\cite{Cook2006}. A~sampling time of $T_s=0.001$, creates the discrete-time state space~matrices

	\[{A} =
	\begin{bmatrix*}[r]
	0.9998 &  -0.0002 &  -0.0108 &   0.0097  & -0.0013\\
	-0.0015 &   0.9789 &   0.0074 &  0  &  0\\
	0.0003  &  0.0037   & 0.9979  &  0  & 0\\
	0 &   0.0010  &  0  &  1.0000 &   0\\
	0  &  0  &  0.0010   & 0 &  1.0000
	\end{bmatrix*}, \quad  
	{B} =
	\begin{bmatrix*}[r]
	0\\
	0.0036\\
	0.0004\\
	0\\
	0
	\end{bmatrix*}.\] 

The learning  parameters for the adaptive learning algorithms are given by $\eta_{a}=\eta_{c}=\alpha_{a}=\alpha_{c}=0.0001$. The~learning parameters are selected to be comparable to the sampling time to have smooth adjustments for the adapted weights. Later, random learning rates are superimposed at each evaluation~step.

The initial conditions are set to
$X_0=
\begin{bmatrix*}[r]
40  &  1.6  &  0.8  & -0.8 &   0.2
\end{bmatrix*}^T.
$

The weighting matrices of the cost functions $D(\dots)$ and $E(\dots)$ are  selected in such a way as to normalize the effects of the different variables in order to increase the sensitivity of the proposed approach against the variations in the measured variables. These are given by
$S=0.0001 \, I_{3 \times 3}$, $M=0.0001$, $R=907$, 
$Q=
\begin{bmatrix*}[c]
0.0625 &   0    &     0   &  0   &      0\\
0  &       25   &     0   &  0   &   0\\
0  &       0    &    25   &  0   &   0\\
0  &       0    &     0   & 100  &   0\\
0  &       0    &     0   &  0   & 100
\end{bmatrix*}.
$

The desired roll-tracking trajectory consists of two smooth opposite turns represented by a sinusoidal reference signal such that $\phi^{desired}(t)=25 \, \sin (2 \, \pi \, t \, /10) \, \deg$ (i.e., right and left turns with max amplitudes of $25 \, \deg$).

\subsection{Simulation~Outcomes}

The simulation scenarios tackle the performance of the standalone tracker, then the characteristics of the overall or combined adaptive control approach. Finally, a~third scenario is considered to discuss the performance of the adaptive learning algorithms under unstructured dynamical environment and uncertain learning parameters. These simulation cases can be detailed out as~follows
\begin{enumerate}[leftmargin=*,labelsep=5mm]
\item Standalone tracker:
The adaptive learning algorithms are tested to achieve only the trajectory-tracking objective (i.e., no overall dynamical optimization is included, and~they are denoted by STA1 and STA2 for Algorithms \ref{alg:alg1} and \ref{alg:alg2} respectively). In~the standalone tracking operation mode, Bellman equations concerning the optimized overall performance and hence the associated optimal control strategies are omitted form the overall adaptive learning structure.
\item Combined control scheme:
This case combines the adaptive tracking control and optimizer schemes (i.e., the tracking control objective is considered along with the overall dynamical optimization using Algorithms \ref{alg:alg1} and \ref{alg:alg2} which are referred to as OTA1 and OTA2 respectively).  
\item Operation under uncertain dynamical and learning environments: The proposed online reinforcement learning approaches are validated using challenging dynamical environment, where the dynamics of the aircraft  (i.e., matrices $A$ and $B$) are allowed to variate at each evaluation step by $\pm50 \%$ around their nominal values at a normal trim condition. The~aircraft is allowed to follow a complicated trajectory to highlight the capabilities of the adaptive learning processes using this maneuver. Additionally, the~actor-critic learning rates are allowed to variate at each iteration index or solution step.    
\end{enumerate}

\subsubsection{Adaptation of the Actor-Critic~Weights}
\begin{figure}[H]
	\centering
	\includegraphics[width=1\linewidth]{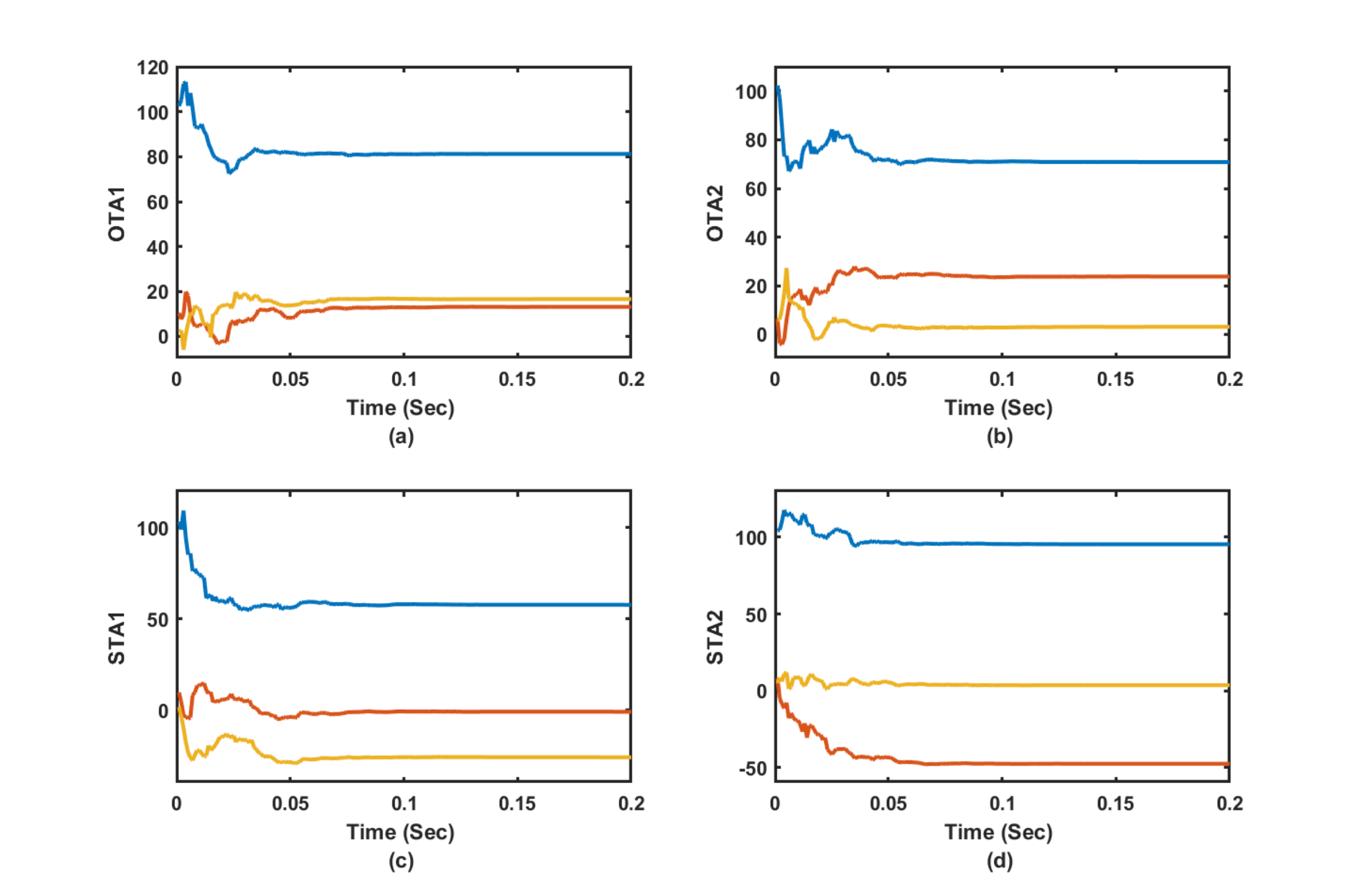}
	\caption{Tuning of the actor weights (associated with the tracking error vector $E$) (\textbf{a}) $\Omega_a$ using OTA1; (\textbf{b}) $\Lambda_a$ using OTA2; (\textbf{c}) $\Omega_a$ using STA1; (\textbf{d}) $\Lambda_a$ using~STA2.}
	\label{fig:actpid}
\end{figure}

\begin{figure}[H]
	\centering
	\includegraphics[width=0.9\linewidth]{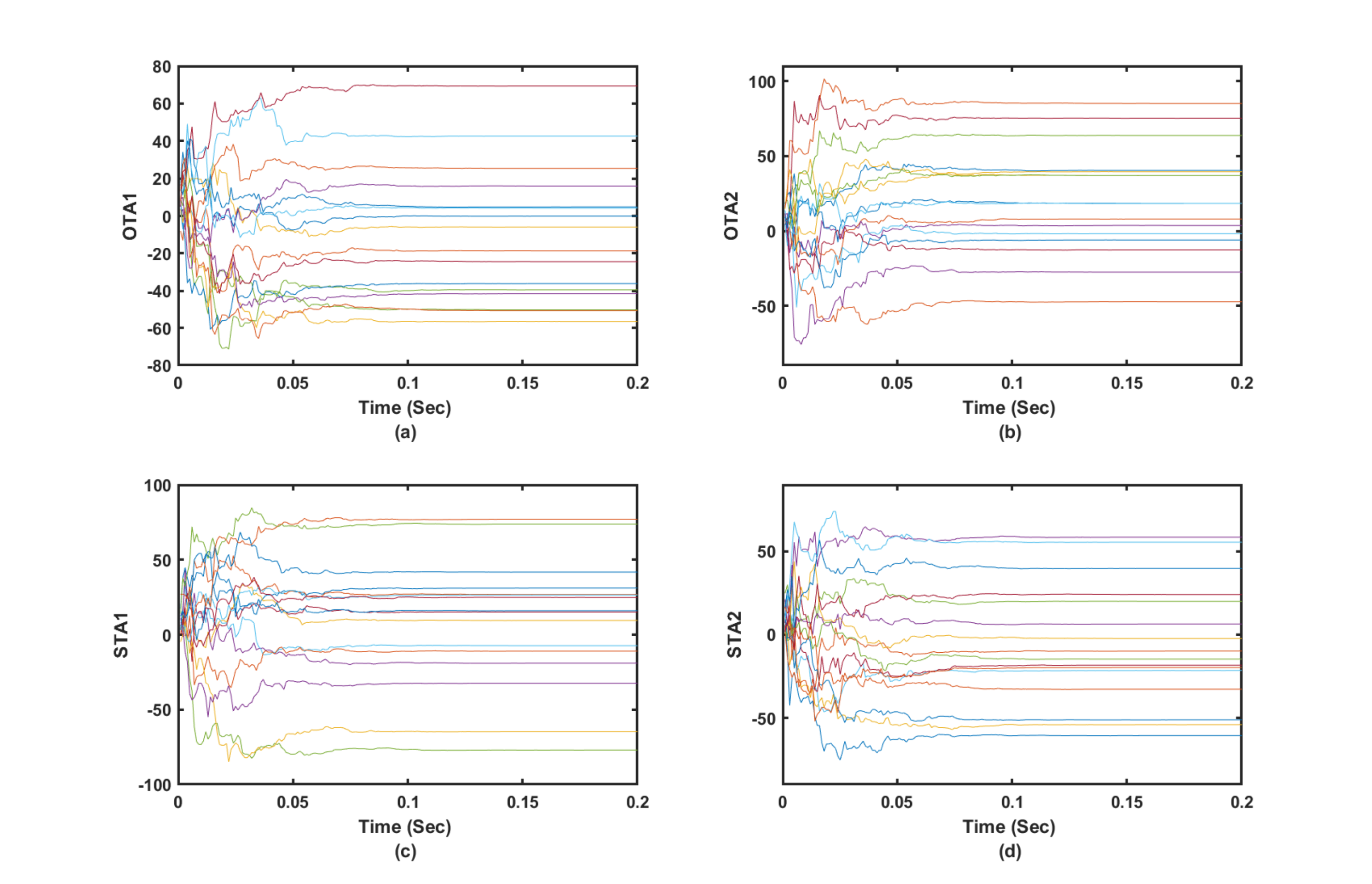}
	\caption{Tuning of the critic weights (associated with the tracking error vector $E$) (\textbf{a}) $\Omega_c$ using OTA1; (\textbf{b}) $\Lambda_c$ using OTA2; (\textbf{c}) $\Omega_c$ using STA1; (\textbf{d}) $\Lambda_c$ using~STA2.}
	\label{fig:critpid}
\end{figure}

\begin{figure}[H]
	\centering
	\includegraphics[width=0.9\linewidth]{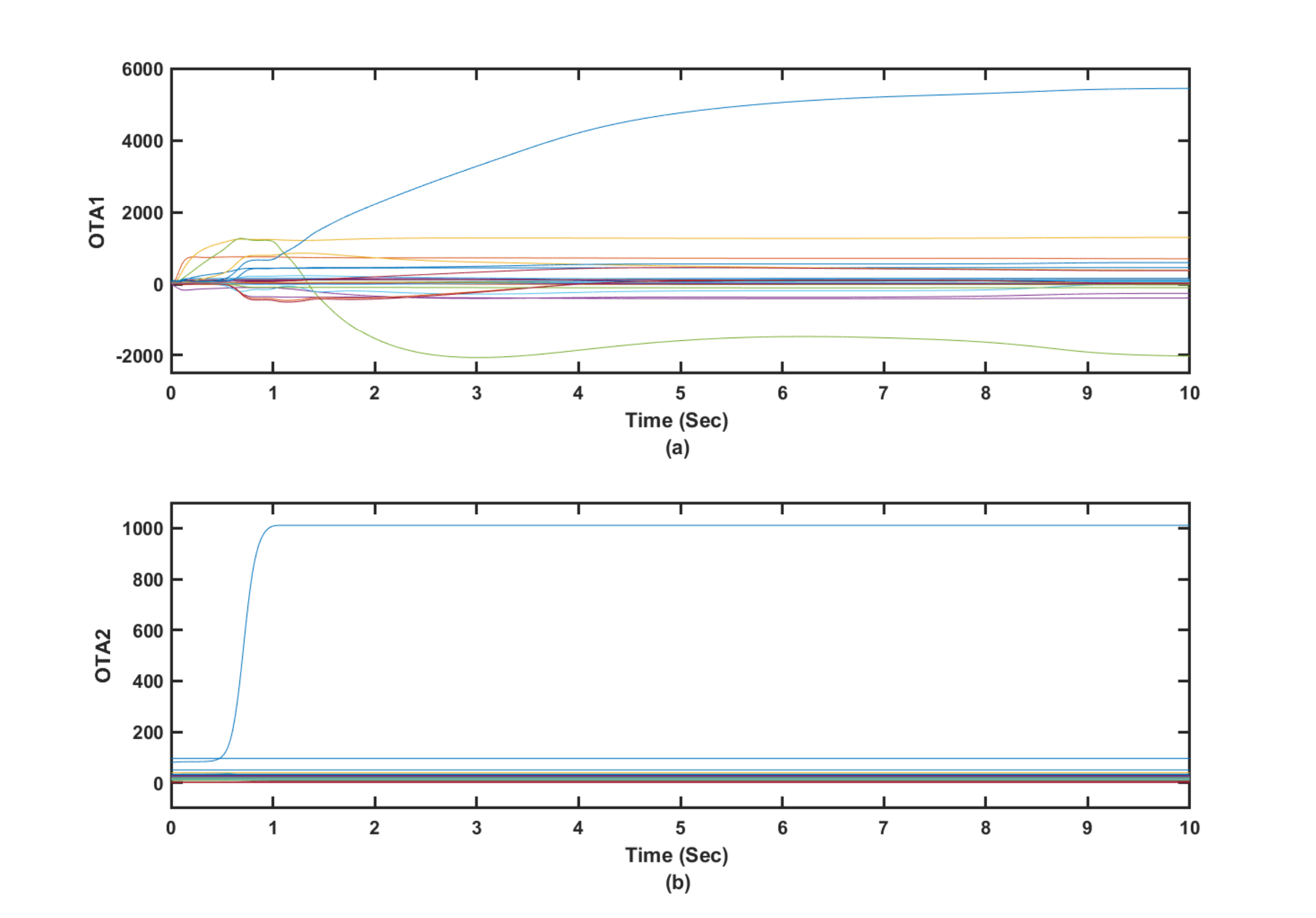}
	\caption{Tuning of the critic weights (associated with the dynamical vector $X$) (\textbf{a}) $\Upsilon_c$ using OTA1; (\textbf{b})~$\Delta_c$ using~OTA2.}
	\label{fig:crtx}
\end{figure}

The tuning processes of the actor and critic weights are shown to converge when they follow solution Algorithms \ref{alg:alg1} and \ref{alg:alg2} as shown in Figures~\ref{fig:actpid}--\ref{fig:crtx}. This is noticed when the tracker is used in a standalone situation or when it is operated within the combined or overall dynamical optimizer. It~is shown that the~actor and critic weights for the tracking component of the optimization process converge in less than $0.1$ s as shown in Figures~\ref{fig:actpid} and \ref{fig:critpid}. The~tuning of the critic weights in the case of optimized tracker took longer time due to the number of involved states and the objective of the overall dynamical optimization problem as shown in Figure~\ref{fig:crtx}. It is worth noting that the tracker part of the controller uses the tracking error signals as inputs which facilitates the tracking optimization process. These results highlight the capability of the adaptive learning algorithms to converge in~real time.

\unskip

\unskip

\subsubsection{Stability and Tracking Error~Measures}
\begin{figure}[H]
	\centering
	\includegraphics[width=0.85\linewidth]{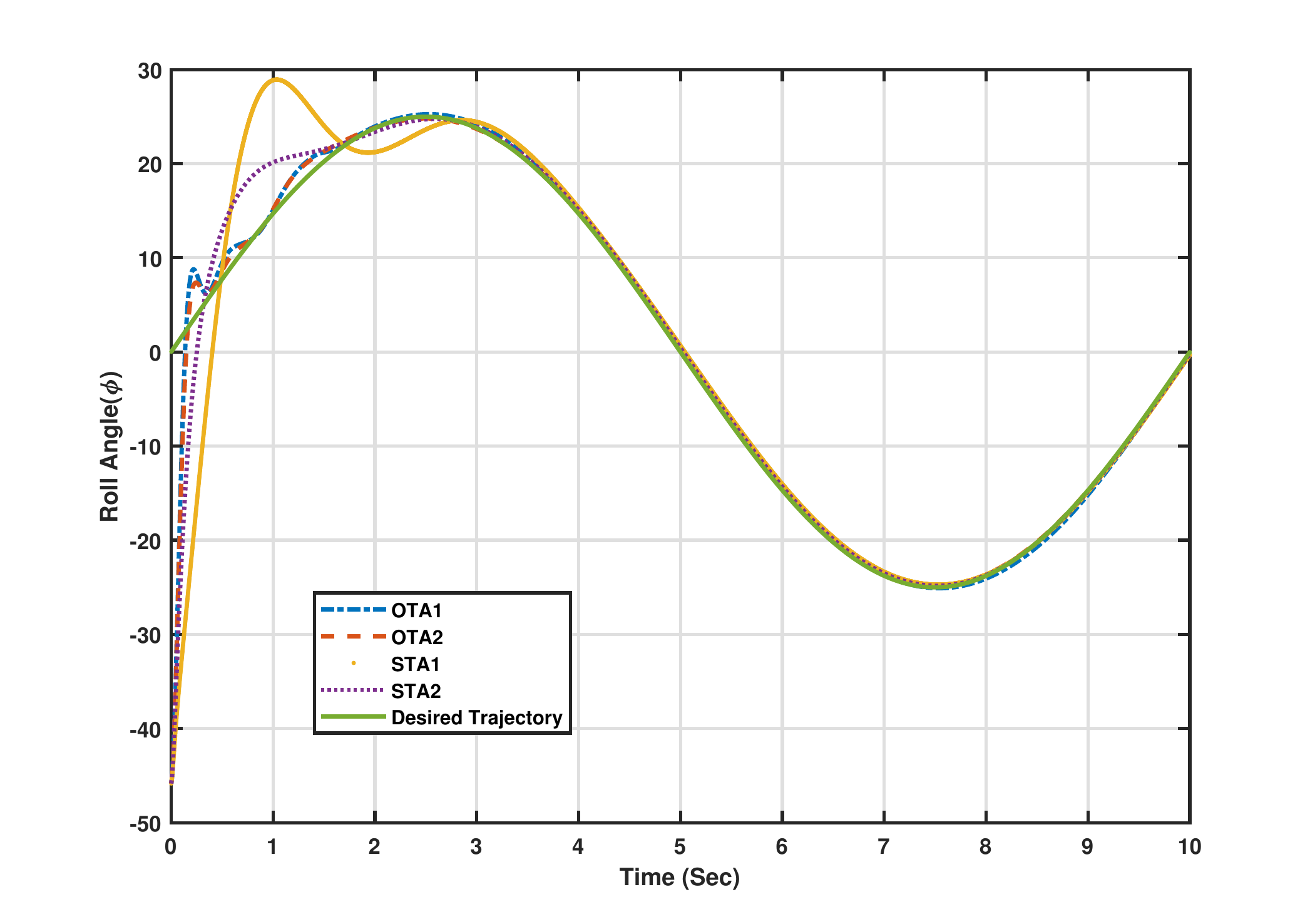}
	\caption{The roll-trajectory-tracking in $\deg$ using OTA1, OTA2, STA1, and~STA2.}
	\label{fig:roll}
\end{figure}

\begin{figure}[H]
	\centering
	\includegraphics[width=0.9\linewidth]{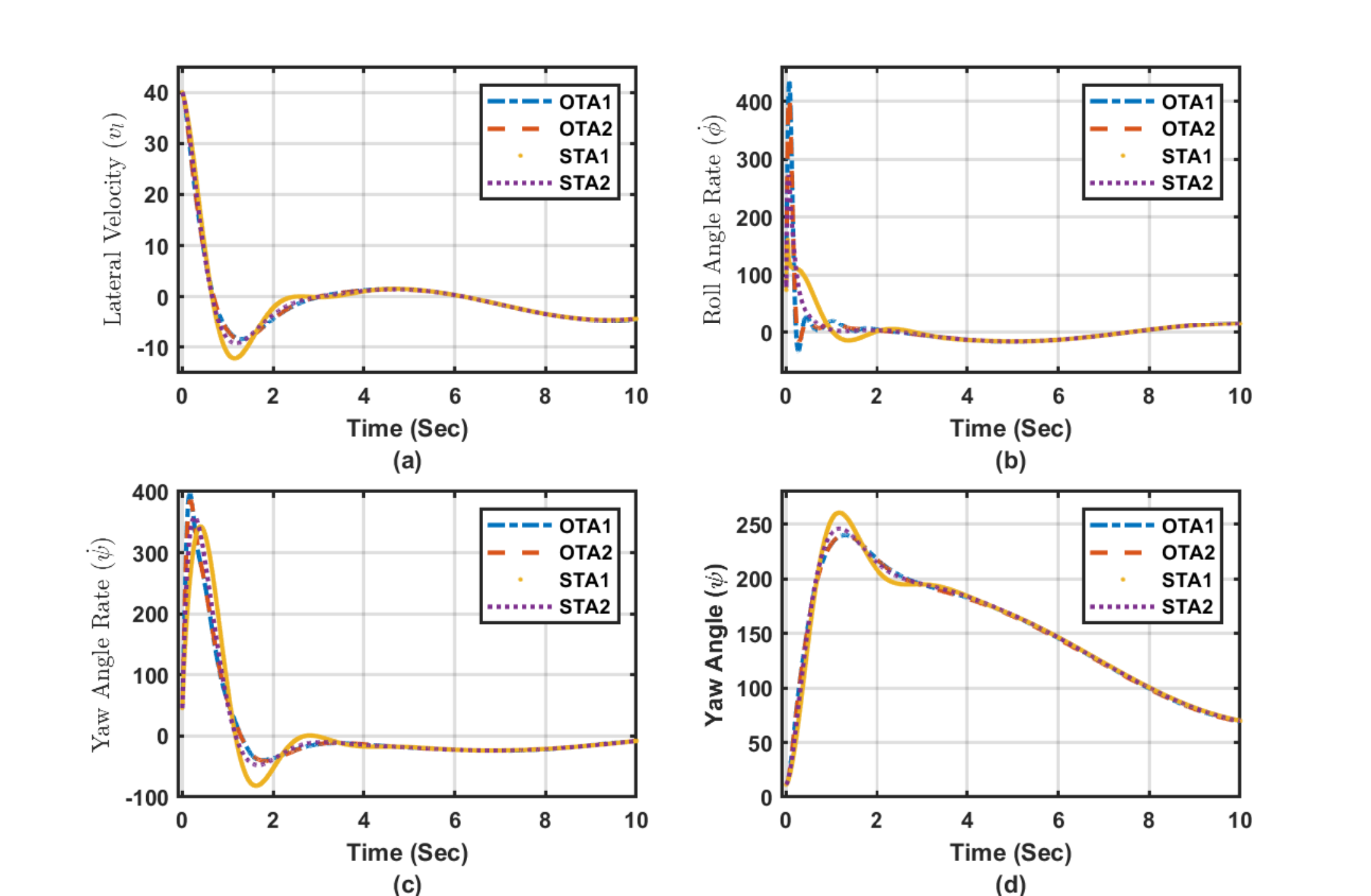}
	\caption{The remaining dynamics using OTA1, OTA2, STA1, and~STA2 (\textbf{a}) Lateral velocity $v_l$ $(m/ \sec)$; (\textbf{b}) Roll angle rate $\dot \phi$ $(\deg/ \sec^2)$; (\textbf{c}) Yaw angle rate $\dot \psi$ $(\deg/ \sec^2)$; and~(\textbf{d}) Yaw angle $\psi$ $(\deg)$.}
	\label{fig:dyn}
\end{figure}

\begin{figure}[H]
	\centering
	\includegraphics[width=0.8\linewidth]{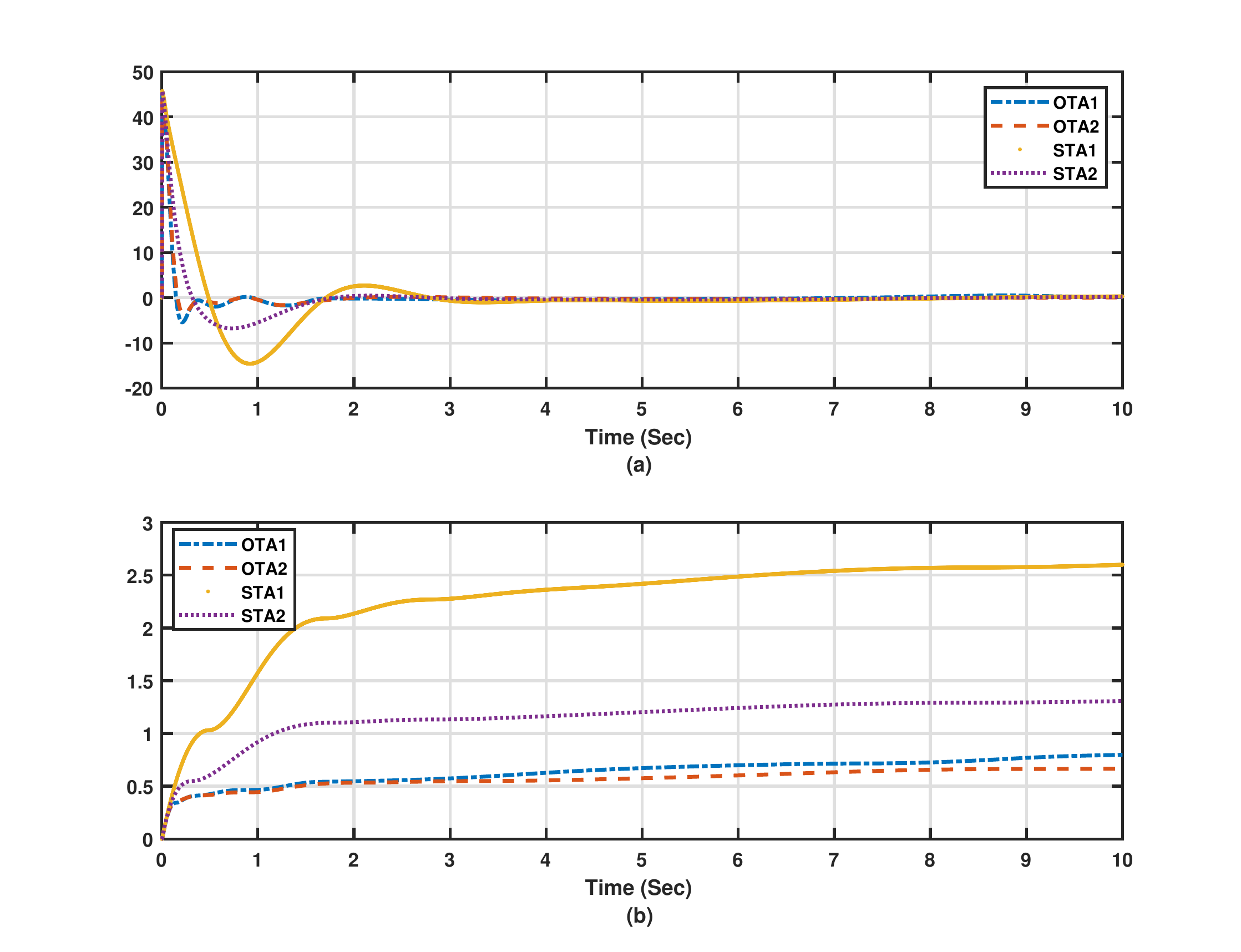}
	\caption{(\textbf{a}) The tracking error signals using OTA1, OTA2, STA1, and~STA2; (\textbf{b}) The average of the accumulated sum of the squared error signals using OTA1, OTA2, STA1, and~STA2.}
	\label{fig:track}
\end{figure}

The adaptive learning algorithms under different scenarios or modes of operation, stabilize the flexible wing system along the desired trajectory as shown in Figures~\ref{fig:roll} and \ref{fig:dyn}. The~lateral motion dynamics eventually follow the desired trajectory. In~this case, the~lateral variables are not supposed to decay to zero, since the aircraft is following a desired trajectory. The~tracking scheme leads this process side by side with the overall energy optimization process, which actually improves the closed-loop characteristics of the aircraft towards minimal energy behavior. It is noticed that Algorithm~\ref{alg:alg2} outperforms Algorithm~\ref{alg:alg1} under standalone tracking mode or the overall optimized tracking mode. In~order to quantify these effects numerically and graphically, the~average accumulated tracking errors obtained using the proposed adaptive learning algorithms are shown in Figure~\ref{fig:track}a,b respectively. These indicate that the optimized tracker modes of operation (i.e., OTA1 and OTA2) give lower errors compared to those achieved during the standalone modes of operations (i.e., STA1~and STA2),{ emphasizing the importance of adding the overall optimization scheme to the tracking system.} Adaptive learning Algorithm~\ref{alg:alg2}, using the optimized tracking mode, achieves the lowest average of accumulated errors as shown in Figure~\ref{fig:track}b. An~additional measure index is used, where the overall normalized dynamical effects are evaluated using the following Normalized Accumulated Cost Index~(NACI)

$$\text{NACI}=\frac{1}{N}\sum_{k=0}^{10 \sec}  [X_{k} ^T \,\,\,\,   U_{Tk}^T]\,\,\, \begin{bmatrix*}[c]
V_1 & 0\\0 & V_2 
\end{bmatrix*} \,\,\, 
\left[\begin{array}{l}
X_{k}\\
U_{Tk}
\end{array}\right],$$ 
where 
$V_1=\begin{bmatrix*}[c]
0.0006    &     0    &     0    &     0   &      0\\
0  &  0.0174    &     0    &     0   &      0\\
0    &     0  &  0.0208    &     0     &    0\\
0    &     0  &       0 &   1.5625     &    0\\
0    &     0  &       0 &        0  &  0.0483
\end{bmatrix*},$
$V_2=0.2268,$ and $N=10,000$  (i.e., the number of iterations during $10$ s) is the total number of~samples.

\unskip
\unskip

\begin{figure}[H]
	\centering
	\includegraphics[width=0.85\linewidth]{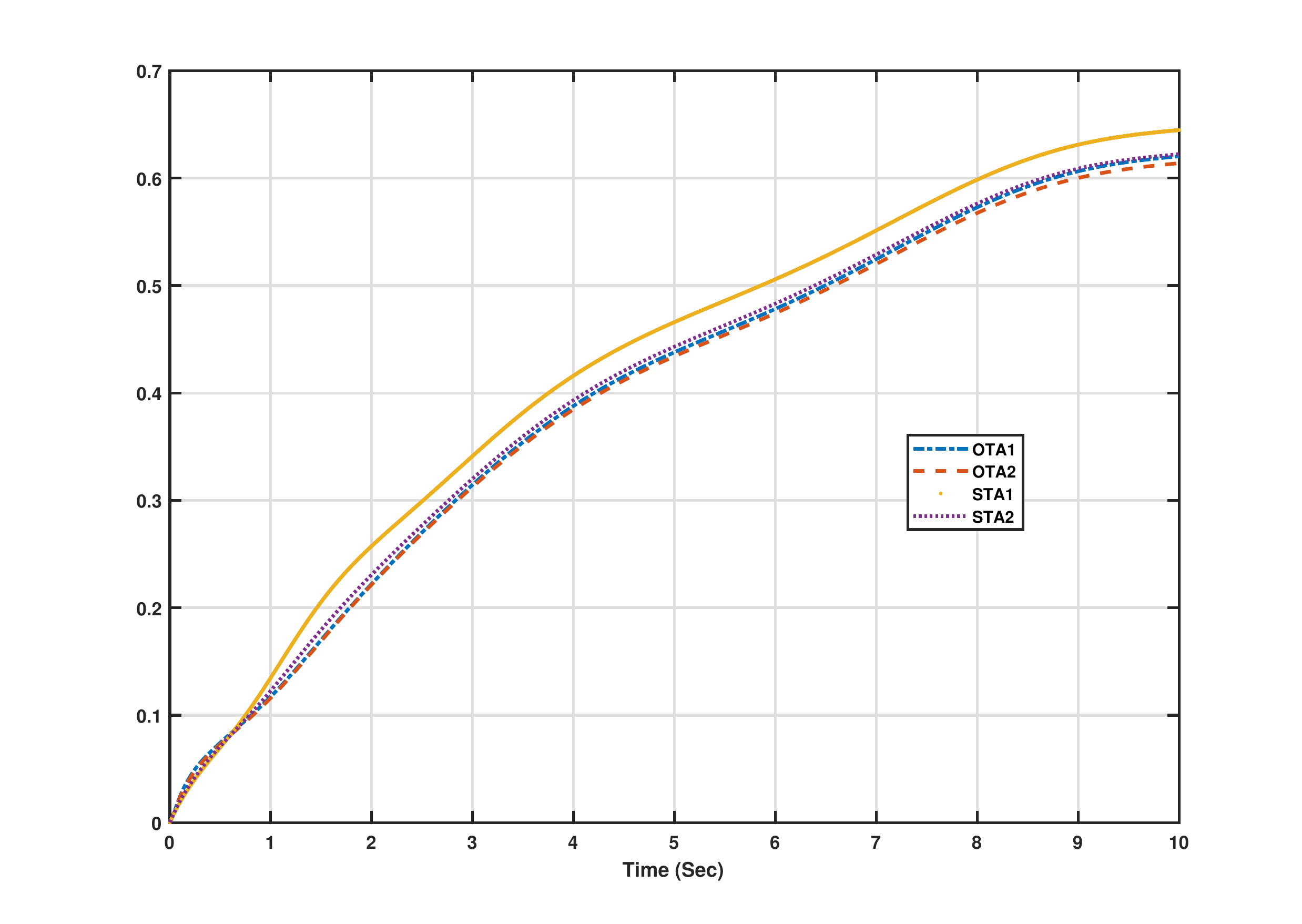}
	\caption{The average of total normalized accumulated dynamical cost using OTA1, OTA2, STA1, and~STA2.}
	\label{fig:totalcost}
\end{figure}

The~normalization values are the square of the maximum measured values of $X_k$ and $U_{Tk}$.
The~adaptive algorithm (OTA2) achieves the lowest overall dynamical cost or effort as shown by Figure~\ref{fig:totalcost}. The~final control laws achieved by using the different algorithms under the above modes of operation (i.e., STA1, STA2, OTA1, and~OTA2) are listed in Table~\ref{tab:Control}.

\begin{figure}[H]
	\centering
	\includegraphics[width=0.85\linewidth]{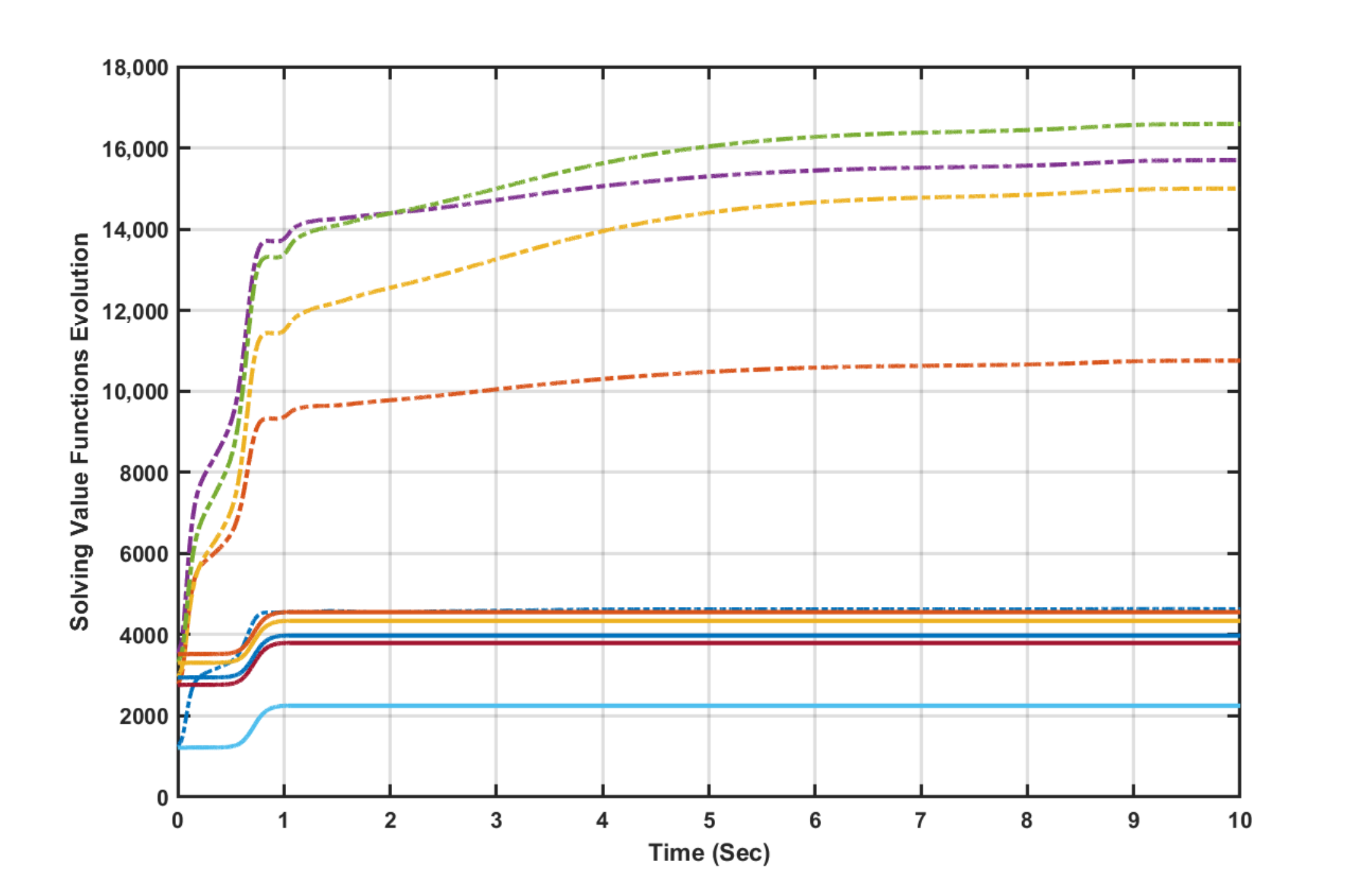}
	\caption{The evolution of the solving value functions $\Gamma^r(\dots), \forall r$ using (OTA1: dashed lines) and (OTA2: solid lines) for five random initial~conditions.
}
	\label{fig:valfunevol}
\end{figure}

The online value iteration processes result in increasing bounded sequences of the solving value functions $\Gamma^r(\dots)$ and $\Xi^r(\dots),\forall r$, which is aligned with the convergence properties of typical value iteration mechanisms. The~online learning outcomes of the value iteration processes $\Gamma^r(\dots)$ (i.e.,~using~Algorithms~\ref{alg:alg1} and \ref{alg:alg2}) are applied and used for five random initial conditions as shown by Figure~\ref{fig:valfunevol}. The~initial solving value functions evaluated by Algorithms \ref{alg:alg1} and \ref{alg:alg2} start from the same positions using the same vector of initial conditions. It is observed that Algorithm~\ref{alg:alg2} (solid lines) outperforms Algorithm~\ref{alg:alg1} (dashed lines) in terms of the updated solving value function obtained using the attempted random initial conditions. Despite both algorithms show general increasing and converging evolution pattern of the solving value functions, value iteration Algorithm~\ref{alg:alg2} exhibits rapid increment and quicker settlement to lower values compared to Algorithm~\ref{alg:alg1}.

\begin{table}[H]
	\centering
	\caption{Final control~laws.}
	\label{tab:Control}\scalebox{0.9}[0.9]{
		\begin{tabular}{ll}
			\toprule
			{\textbf{Method}} & \textbf{Control Law} \\
			\midrule
			\multirow{1}{*}{$\Omega_a$ (STA1)} & $ 
			\begin{bmatrix*}[r]
			57.5021 &  -1.1475 & -26.1183
			\end{bmatrix*}
			$ \\ 
			\midrule
			\multirow{1}{*}{$\Lambda_a$ (STA2)} & $ 
			\begin{bmatrix*}[r]
			95.3475 & -47.6060  &  3.5581
			\end{bmatrix*}
			$ \\ 
			\midrule
			\multirow{1}{*}{$\Omega_a$ (OTA1)} & $ 
			\begin{bmatrix*}[r]
			81.2142 &  13.2197 &  16.6757
			\end{bmatrix*}
			$ \\
			\midrule
			\multirow{1}{*}{$\Lambda_a$ (OTA2)} & $ 
			\begin{bmatrix*}[r]
			70.8768 &  23.9006  &  3.0130
			\end{bmatrix*}
			$ \\
			\midrule
			\multirow{1}{*}{$\Upsilon_a$ (OTA1)} & $ 
			\begin{bmatrix*}[r]
			-0.0535 &  -0.0897 &  -0.1386 &  -0.3704 &  -0.3545
			\end{bmatrix*}
			$ \\
			\midrule
			\multirow{1}{*}{$\Delta_a$ (OTA2)} & $ 
			\begin{bmatrix*}[r]
			-0.0422 &  -0.1487 &  -0.3479 &  -0.4356 &  -0.1217
			\end{bmatrix*}
			$ \\
			\bottomrule
	\end{tabular}}
\end{table}
\unskip

\unskip

\subsubsection{Closed-Loop~Characteristics}
To examine the time-characteristics of the adaptive learning algorithms, the~closed-loop performances of the adaptive learning algorithms under the optimized tracking operation mode (i.e.,~OTA1 and OTA2) are plotted in Figure~\ref{fig:eigv}. 
Apart from the tracking feedback control laws, the~optimizer state feedback control laws directly affect the closed-loop system. The~forthcoming analysis is to show how (1) the aircraft system initially starts (i.e., open-loop system); (2) the evolution of the closed-loop poles during the learning process; and~(3) the final closed-loop characteristics when the actor weights finally converge.
The trace of the closed-loop poles achieved using OTA2 (\mbox{i.e., the {\large \color{blue} $\boldsymbol{\bullet}$} marks}) shows concise and faster stable behavior than that obtained using OTA1  (\mbox{i.e., the {\large \color{red} $\boldsymbol{\bullet}$} indicators}), and~definitely faster than the open-loop characteristics. The~dominant open-loop pole is moved further into the stability region, when the overall dynamical optimizer is included, as~listed in Table~\ref{tab:eigvw}. These results emphasize the stability and superior time-response characteristics achieved using the adaptive learning approaches, especially Algorithm~\ref{alg:alg2}.

\begin{table}[H]
	\centering
	\caption{Open- and closed-loop~eigenvalues.}
	\label{tab:eigvw}\scalebox{0.9}[0.9]{
		\begin{tabular}{ll}
			\toprule
			{\textbf{Method}} & \textbf{Poles} \\
			\midrule
			\multirow{1}{*}{Open-loop system} \vspace{5pt}& $0, \quad -0.2752 \pm 0.8834i,$ \\ 
			(STA1 and STA2) & $-0.5088, \quad -22.5902$\\
			\midrule
			\multirow{1}{*}{Closed-loop system} \vspace{5pt}& $-0.0169 \pm 0.9393i,$ \\ 
			\multirow{1}{*}{(OTA1)}& $-0.3771, \quad -0.8736, \quad -22.7489$\\
			\midrule
			\multirow{1}{*}{Closed-loop system} \vspace{5pt}& $-0.0768 \pm 0.9409i,$ \\
			\multirow{1}{*}{(OTA2) }& $-0.1152, \quad -1.2600, \quad -22.8079$\\ 
			\bottomrule
	\end{tabular}}
\end{table}

\subsubsection{Performance in Uncertain Dynamical~Environment}
\begin{figure}[H]
	\centering
	\includegraphics[width=0.85\linewidth]{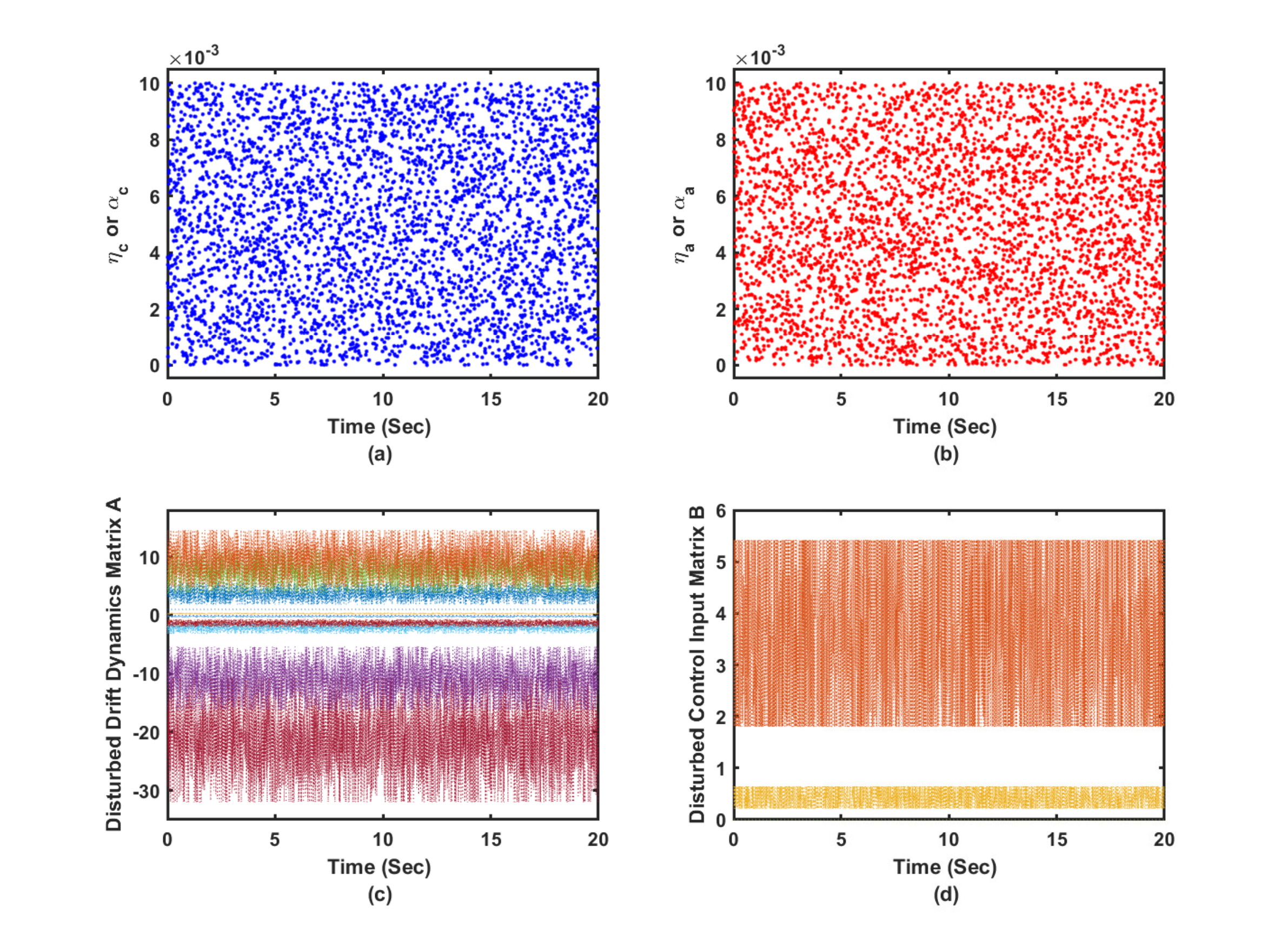}
	\caption{Variations in the dynamical learning environment (\textbf{a}) Variations in the critic learning rates $\eta_c=\alpha_c$; (\textbf{b}) Variations in the actor learning rates $\eta_a=\alpha_a$; (\textbf{c}) Uncertainties in the entries of the drift dynamics matrix $A$; and~(\textbf{d}) Uncertainties in the entries of the control input matrix $B$.}
	\label{fig:rev_variation}
\end{figure}

This simulation scenario challenges the performance of the online adaptive controller in uncertain dynamical environment. The~continuous-time aircraft aerodynamic model (i.e., the aircraft state space model with the drift dynamics matrix $A$ and control input matrix $B$) is forced to involve unstructured dynamics~\cite{Cook2006}. These disturbances are of amplitudes $\pm 50\%$ around the nominal values at the trim condition and they are generated from a normal Gaussian distribution as shown in Figure~\ref{fig:rev_variation}c,d. Additionally, the~sampling time is set to $T_s=0.005$ s and the actor-critic learning rates are allowed to vary at each evaluation step as shown by Figure~\ref{fig:rev_variation}a,b to test a band of learning parameters. Finally, a~challenging desired trajectory is proposed such that $\phi^{desired}(t)=(25 \, \sin (6 \, \pi \, t \, /10) \,+15 \, \cos (16 \, \pi \, t \, /10)) \,  e^{-3\, t/10}  \deg$. These coexisting factors challenge the effectiveness of the controller. The~randomness which appears in the proposed coexisting dynamical learning situations provides rich exploration environment for the adaptive learning processes. These dynamic variations occur at each evaluation step which guarantees some sort of generalization for the dynamical processes under~consideration.

\begin{figure}[H]
	\centering
	\includegraphics[width=0.8\linewidth]{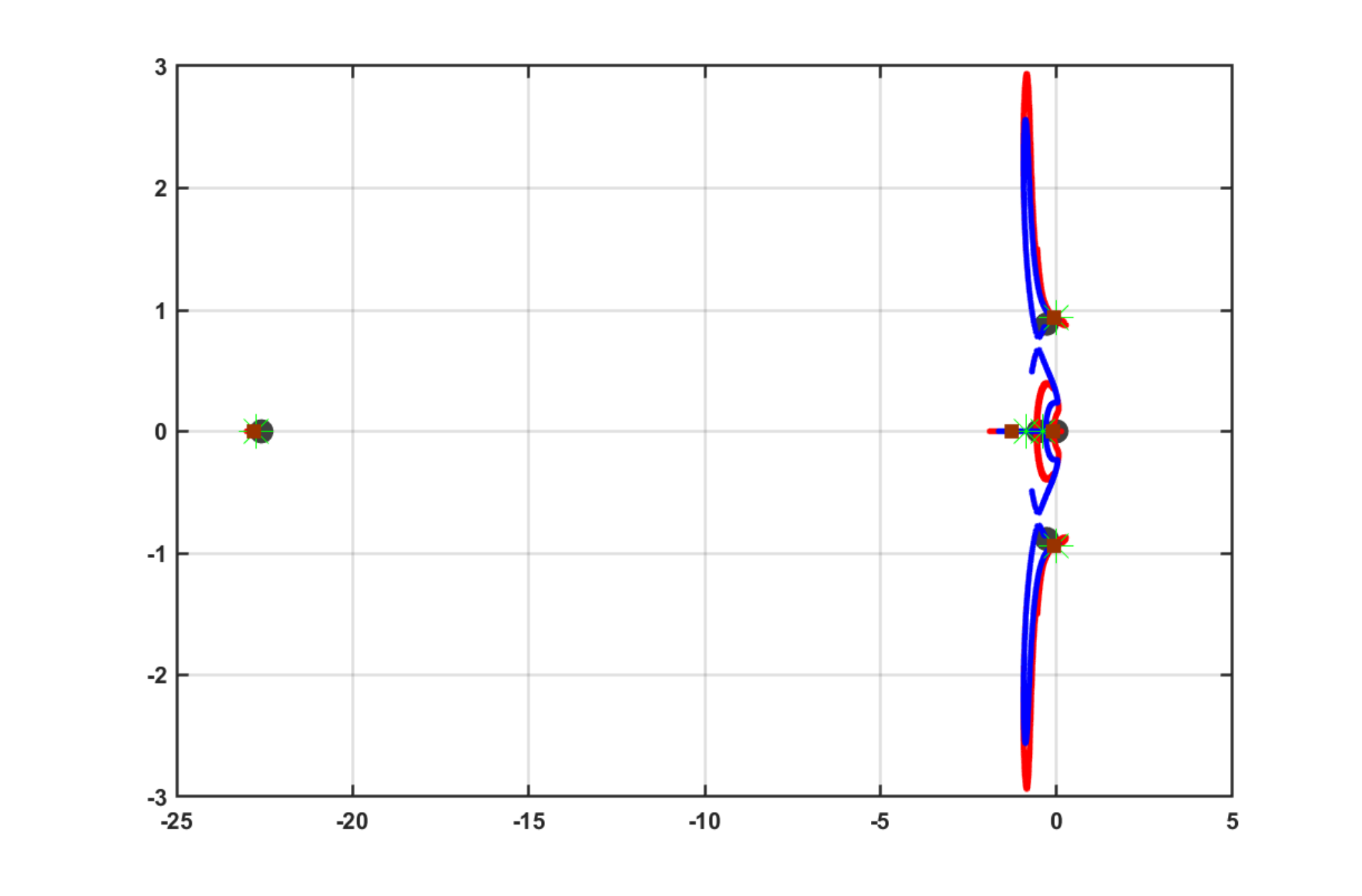}
	\caption{The closed-loop poles of the flexible wing system using OTA1 and OTA2.
The notations {\color{darkbrown} $\CIRCLE$} refer to the open-loop poles of the system. The~closed-loop poles during the online learning process evaluated by OTA1 and OTA2 are remarked by {\large \color{red} $\boldsymbol{\bullet}$} and {\large \color{blue} $\boldsymbol{\bullet}$} symbols respectively. The~final closed-loop poles using OTA1 are denoted by the $\Large \color{green} \pmstar$ marks, while those obtained by OTA2 are given the $ \crule[darkbrown]{.2cm}{.2cm}$ notations.}
	\label{fig:eigv}
\end{figure}

\begin{figure}[H]
	\centering
	\vspace{6pt}
	\includegraphics[width=0.85\linewidth]{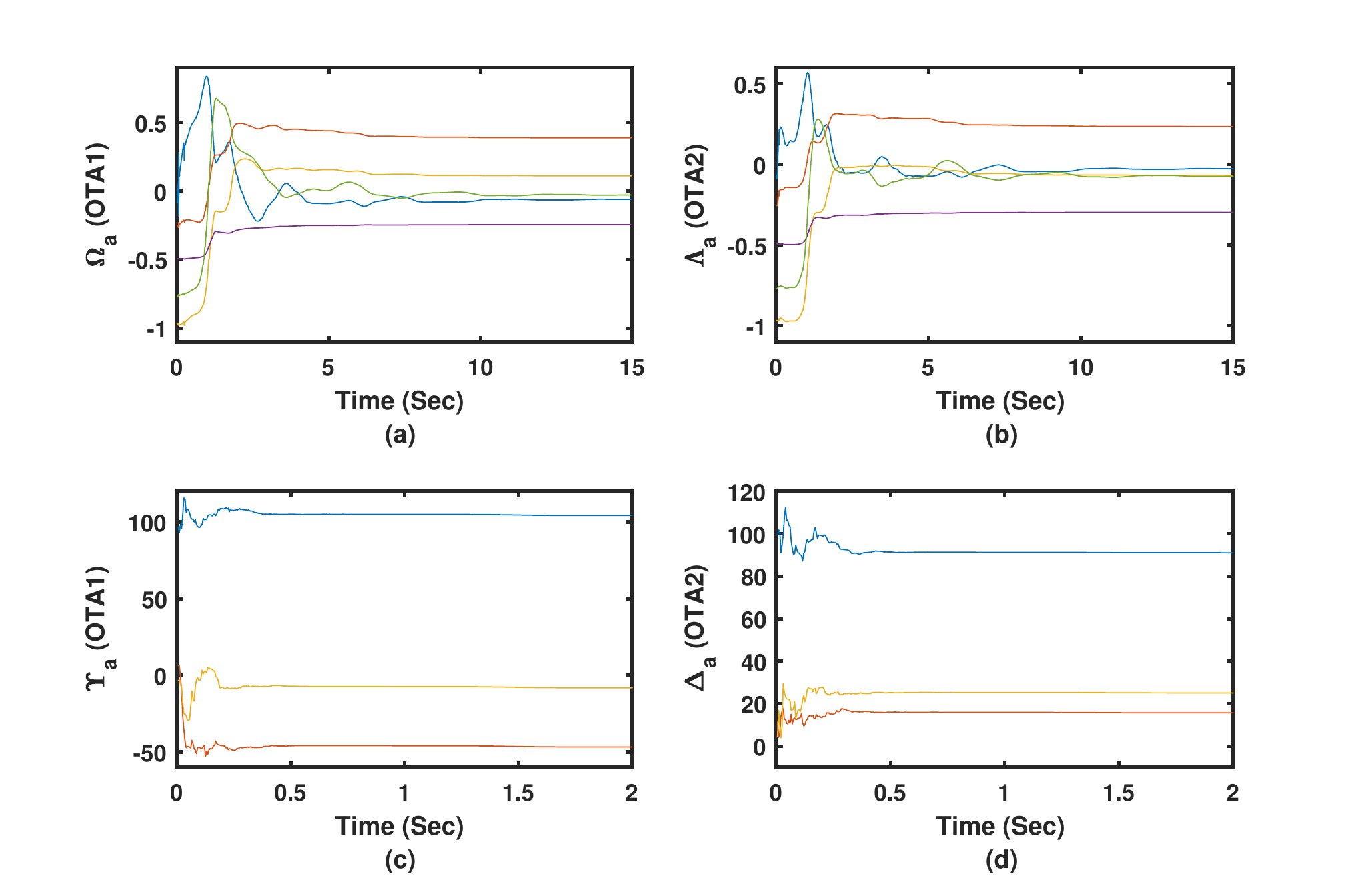}
	\caption{Tuning of the actor weights (\textbf{a}) $\Omega_a$ using OTA1; (\textbf{b}) $\Lambda_a$ using OTA2; (\textbf{c}) $\Upsilon_a$ using OTA1; (\textbf{d})~$\Upsilon_a$ using~OTA2.
}
	\label{fig:rev_actor}
\end{figure}

\begin{figure}[H]
	\centering
	\includegraphics[width=0.9\linewidth]{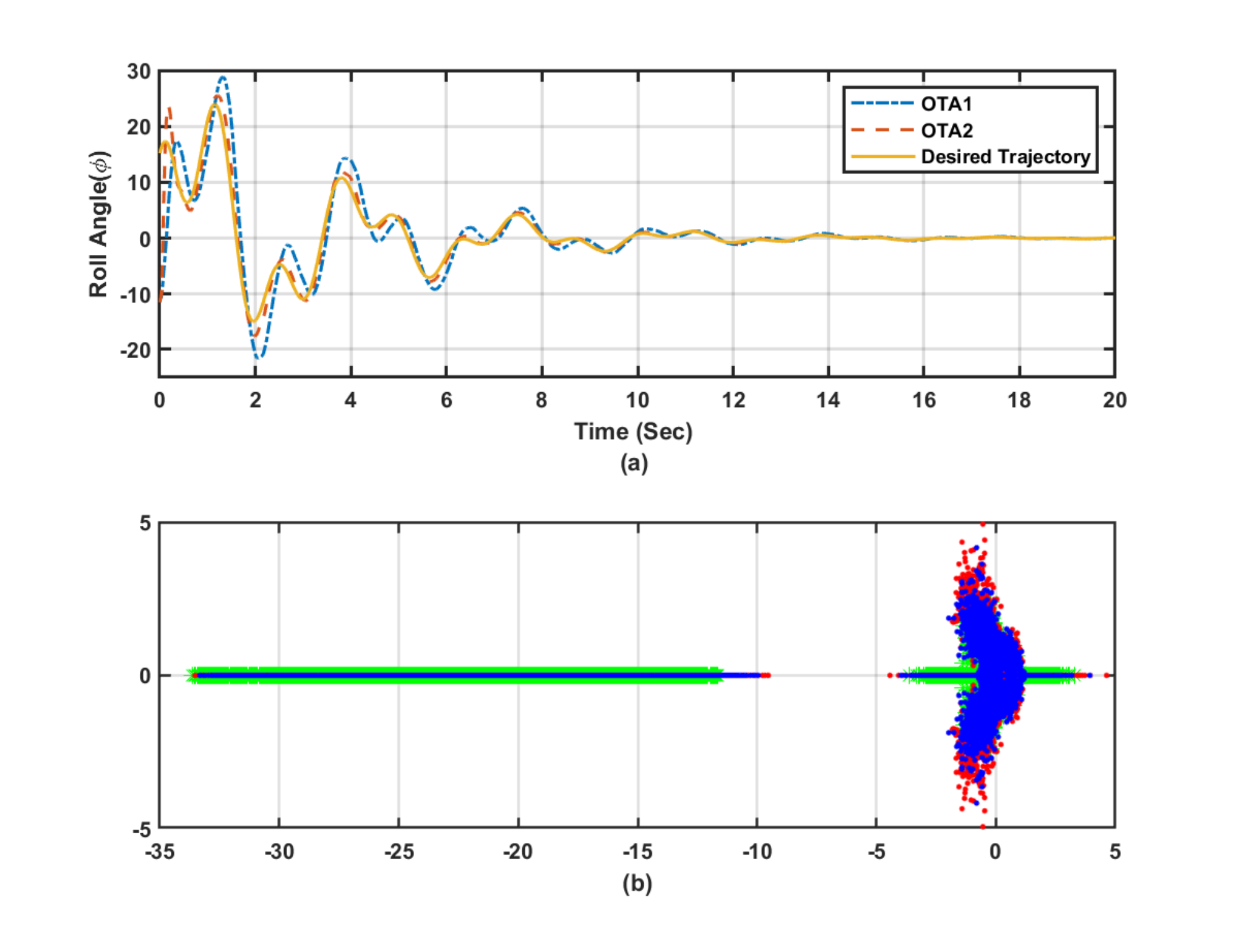}
	\caption{The performance in uncertain dynamical environment (\textbf{a}) The roll-trajectory-tracking in $\deg$ using OTA1 and OTA2; (\textbf{b}) The closed-loop poles during the online learning process evaluated by OTA1 and OTA2 are remarked by {\large \color{red} $\boldsymbol{\bullet}$} and {\large \color{blue} $\boldsymbol{\bullet}$} symbols respectively. The~open-loop poles of the disturbed dynamical system are denoted by the $\Large \color{green} \pmstar$ marks.}
	\label{fig:rev_dyn_eig}
\end{figure}

Figure~\ref{fig:rev_actor}a--d emphasize that the adaptive learning Algorithms~\ref{alg:alg1} and~\ref{alg:alg2} (i.e., OTA1 and OTA2) are able to achieve the trajectory-tracking objectives. The~actor weights are shown to successfully converge despite the co-occurring uncertainties. The~adaptation processes are effectively responding to the acting disturbances, where relatively longer time is needed to converge to the proper control gains. The~tracking feedback control gains took shorter time to converge as shown by Figure~\ref{fig:rev_actor}c,d, where the tracking feedback control law depends only on the state $\phi_k$, and~implicitly its derivative. Algorithm~\ref{alg:alg2} exhibited better trajectory-tracking features compared to those obtained using Algorithm~\ref{alg:alg1} as shown by Figure~\ref{fig:rev_dyn_eig}a. Figure~\ref{fig:rev_dyn_eig}b, when compared to Figure~\ref{fig:eigv}, shows how the open-loop poles, represented by $\Large \color{green} \pmstar$ marks (recorded disturbances at each iteration \textit{k}), spread all over the S-plane. The~adaptive learning Algorithms~\ref{alg:alg1}~and~\ref{alg:alg2}, exhibited similar stable behavior as observed in the earlier scenarios. However, longer time was needed to reach asymptotic stability around the desired reference trajectory. This can be observed by examining the spread of the closed-loop poles obtained using OTA1 ({\large \color{red} $\boldsymbol{\bullet}$} notations) and OTA2 ({\large \color{blue} $\boldsymbol{\bullet}$} symbols). These results highlight the insensitivity of the proposed adaptive learning approaches against different uncertainties in the dynamic learning~environments.

\unskip

\unskip
\unskip

\section{Implications in Practical Applications and Future Research~Developments}
\label{sec:futurework}

The proposed combined adaptive learning approach can be integrated into various complex robotic or nonlinear system applications using extremely flexible adaptive learning black-box mechanisms.  These are keen to optimize the performance of the actuation devices while maintaining the tracking control mission in an online fashion. At~least, it will enable complicated distributed tracking solutions for structured robotic systems using simple adaptation laws with affordable computational costs compared to existing adaptive approaches. It can work in unstructured dynamical enthronements where it is really difficult to have full dynamical models for the underlying systems. The~proposed adaptive learning algorithms can be deployed directly into the control units, where the only precautions are concerned with; (1) matching the sampling frequency (imposed by the sensory devices) to the learning parameters; (2) conditioning the weighting matrices in the utility or cost functions according to the actuation signals and the measured variables. The~proposed learning approach is adaptive to the selection of the measured states which makes it convenient to use in many real-world applications, since it does not rely on complicated adaptive learning~constraints.

Future research directions may extend other reinforcement learning tools, such as policy iteration schemes, in~order to develop combined adaptive tracking processes. This direction should find means to tackle the admissibility requirements of the initial policies along with relaxing the computational efforts required to accomplish these processes. The~proposed adaptive learning approaches can be adopted for multi-agent applications. Taking into consideration the complexity of the multi-agent structures, this would involve further research investigations which tackle connectivity, communication costs, and~stabilizability of the coupled control schemes as well as the convergence conditions for the adaptive learning solutions. These ideas may consider structures based on Bellman equations as well as the Hamilton–Jacobi–Bellman equations. Additional directions may investigate the use of other approximate dynamic programming classes which employ gradient-based solving forms to solve the optimal tracking control problem~\cite{Sutton_1998,Lewis_2012}. These involve solutions for the Dual Heuristic Dynamic Programming and Action-Dependent Dual Heuristic Dynamic Programming problems.  These developments should handle the dependence of the temporal difference solutions on the complete dynamical model~information.

\section{Conclusions}
\label{sec:Conclusion}
A class of tracking control problems is solved using online model-free reinforcement learning processes. The~formulation of the optimal control problem tackled the tracking as well the overall dynamical processes by formulating the respective Bellman optimality or temporal difference equations. Two separate linear feedback control laws are adapted simultaneously in real time, where the first linear feedback law decides the optimal control gains associated with a flexible tracking error structure and the second law optimizes the overall dynamical performance during the tracking process. The~proposed approach is employed to solve the challenging trajectory-tracking control problem of a flexible wing aircraft, were the aerodynamics of the wing are unknown and difficult to capture in a dynamical model. An~aggressive learning environment that involves complicated reference trajectory, uncertain~dynamical system, and~flexible learning rates is adopted to show the usefulness of the developed learning approach. The~complete optimized tracker revealed better closed-loop characteristics than those obtained using the standalone~tracker.

\vspace{6pt}

\authorcontributions{All authors have made great contributions to the work. Conceptualization, M.A., W.G. and D.S.; Methodology, M.A., W.G. and D.S.; Investigation, M.A.; Validation, W.G. and D.S.;  Writing-Review \& Editing, M.A., W.G. and~D.S.}

\funding{This research was partially funded by Ontario Centers of Excellence (OCE) and the Natural Sciences and Engineering Research Council of Canada (NSERC).}
\conflictsofinterest{The authors declare no conflict of interest. The~funding sponsors had no role in the design of the study; in the collection, analyses, or~interpretation of data; in the writing of the manuscript, and~in the decision to publish the~results.}

\abbreviations{The following abbreviations are used in this manuscript:\\
\noindent 
\begin{tabular}{@{}ll}
{Variables}\\
$v_{l}$ & lateral velocity in the wing's frame of motion.\\
$\phi$, $\psi$ & Roll and yaw angles in the wing's frame of motion.\\
$\dot \phi$, $\dot \psi$& Roll and yaw angle rates in the wing's frame of motion. \\
{Abbreviations}\\
 ADP&Approximate Dynamic Programming \\
 HDP & Heuristic Dynamic Programming\\
 DHP & Dual Heuristic Dynamic Programming\\
 ADHDP & Action-Dependent Heuristic Dynamic Programming\\
 ADDHP & Action-Dependent Dual Heuristic Dynamic Programming\\
 RL & Reinforcement Learning \\
 HJB & Hamilton--Jacobi--Bellman\\
 PD & Proportional-Derivative\\
 PID & Proportional-Integral-Derivative\\
 OTA1 & Optimized Tracking Using Algorithm 1\\
 OTA2 & Optimized Tracking Using Algorithm 2\\
 STA1   & Standalone Tracking Using Algorithm 1\\
 STA2   & Standalone Tracking Using Algorithm 2\\
      
      & 
\end{tabular}}

\reftitle{References}


\end{document}